\documentclass[reprint,floatfix, pre, aps, longbibliography]{revtex4-2}

\usepackage{xfrac}
\usepackage{amsmath}
\usepackage{amssymb}
\usepackage{dsfont}

\usepackage[unicode,psdextra]{hyperref}
\usepackage{graphicx}
\usepackage{xcolor}
\hypersetup{
    colorlinks,
    linkcolor={red!50!black},
    citecolor={blue!50!black},
    urlcolor={blue!90!black}
}

\usepackage{physics}

\usepackage[english]{babel}
\usepackage[T1]{fontenc}

\usepackage{libertine}

\usepackage{framed}
\colorlet{shadecolor}{blue!20}
\DeclareMathOperator{\erfc}{erfc}
\newcommand{\pd}{\partial}
\newcommand{\ii}{\mathrm{i}}

\begin{document}

\title{Occupation time of a system of Brownian particles on the line with steplike initial condition}
\author{Ivan N. Burenev}
\author{Satya N. Majumdar}
\author{Alberto Rosso}
\affiliation{LPTMS, CNRS, Universit\'e Paris-Saclay, 91405 Orsay, France}

\begin{abstract}
We consider a system of noninteracting Brownian particles on the line with steplike initial condition and study the statistics of the occupation time on the positive half-line.
We demonstrate that even at large times, the behavior of the occupation time exhibits long-lasting memory effects of the initialization.
Specifically, we calculate the mean and the variance of the occupation time, demonstrating that the memory effects in the variance are determined by a generalized compressibility (or Fano factor), associated with the initial condition.
In the particular case of the uncorrelated uniform initial condition we conduct a detailed study of two probability distributions of the occupation time: annealed (averaged over all possible initial configurations) and quenched (for a typical configuration). 
We show that at large times both the annealed and the quenched distributions admit 
large deviation form and we compute analytically the associated rate functions. 
We verify our analytical predictions via numerical simulations using Importance Sampling Monte-Carlo strategy. 
\end{abstract}

\maketitle

\section{Introduction}

\par A classical problem in out-of-equilibrium statistical mechanics is the problem of effusion, i.e. the process in which gas of particles leak from the reservoir through the small hole. 
In its simplest one-dimensional realization this process can be modeled by initially confining particles within a box and then removing one of the walls. 
The system is intrinsically out-of-equilibrium which results in nontrivial behaviors even in this seemingly simple case. 

\par One of the peculiar features of such systems lies in their ability to retain an everlasting memory of the initialization. In other words, different initial conditions lead to distinct behaviors and the differences do not fade out with time. 
For diffusive particles this phenomenon has been observed in various observables, such as  total particle current \cite{DG-09,KM-12}; local time at the origin \cite{BMR-23,SB-23}; displacement of a tracer \cite{BJC-22,SD-23}.
Memory effects also persist in more general Gaussian processes \cite{DMS-23} and in other systems like run-and-tumble particles \cite{BMRS-20,JRR-23,JRR-23b}; Brownian particles with resetting \cite{BHMMRS-23,HMS-23} or Jepsen gas~\cite{BM-07}.

\par Here we focus exclusively on noninteracting Brownian particles. In this case an important class of observables is so-called Brownian functionals \cite{M-05-BFPCS}. Denoting by $x_i(t)$ the trajectory of the $i$-th particle, and by $t$ total time of the process, we define these observables by 
\begin{equation}\label{eq:BrownianFunctional=def}
  O \equiv \sum_{i}\int_{0}^{t} V\left[ x_i(t') \right]\,\dd t',
\end{equation}
where $V[x]$ is an arbitrary function. 

\par 
It is important to note, that the probability distribution of the functional for an individual particle is inherently dependent on its initial position, hence \eqref{eq:BrownianFunctional=def} involves a sum of independent yet \textit{nonidentically} distributed random variables. Consequently the general theory well-developed for identically distributed random variables is inapplicable here.  
At the same time, initial condition, as was argued in \cite{DG-09}, plays a role similar to the realization of the disorder in the theory of disordered systems. Thus there are two natural ways to treat it. The first way is to average over all realizations (``annealed'' scheme) and second is to consider a typical initial configuration (``quenched'' scheme).

\par In our previous paper \cite{BMR-23} we conducted a study of a specific functional, local time density at the origin, which corresponds to the choice $V[x] = \delta( x )$ in \eqref{eq:BrownianFunctional=def}. Here we extend the analysis to another observable. Specifically,  we opt for $V[x]$ in \eqref{eq:BrownianFunctional=def} to be a Heaviside function $V[x] = \theta(x)$
\begin{equation}\label{eq:OccupationTime=def}
  T \equiv \sum_{i} \int_{0}^{t} \theta\left[ x_i(t') \right]\, \dd t'.
\end{equation}
This quantity, usually referred to as occupation (or residence) time on the half-line, characterizes the amount of time particles have spent to the right of the origin.

\par The single particle counterpart of \eqref{eq:OccupationTime=def} has been already extensively studied in both mathematics \cite{L-58,K-62,Borodin-Handbook} and physics literature for various systems. A nonexhaustive list of examples includes diffusion in the Sinai-type potential \cite{MC-02-LOTRM,SMC-06}; fractional Brownian motion \cite{SDW-18}; Gaussian Stationary process \cite{SB-02,EMB-04}; run-and-tumble particle \cite{SK-19}; random acceleration model \cite{BBBRZC-16}; renewal processes \cite{GL-00,BB-11}; trap model \cite{BB-07}; continuous-time random walk \cite{BB-05};  coarsening systems \cite{DG-98,NT-98,BBDG-99}; heterogeneous diffusion \cite{S-22}; diffusion with a drift \cite{NT-17,NT-18}; diffusion with stochastic resetting \cite{HMMT-19,SMS-23} and also for a class of Gaussian Markov processes \cite{DM-99}. 

\par
The concept of occupation time arises in very diverse domains. Just to name a few examples, it appears in the context of blinking quantum dots \cite{EHB-09}; problems of interface growth \cite{TNDS-99}, spin glasses \cite{SD-02} and stochastic thermodynamics \cite{BREMP-18}; competitive sports \cite{CKR-15} and finance \cite{L-99}.

\par 
In this paper we study the statistical properties of the occupation time on the positive half-line \eqref{eq:OccupationTime=def} for a system of noninteracting Brownian particles on the line with steplike initial condition. Following the approach of \cite{BJC-22} we demonstrate that the memory effects in the variance of the occupation time are governed by the generalized compressibility (or Fano factor) of the initial condition. 
For the particular case of uncorrelated uniform initial condition we study the tails of both quenched and annealed probability distributions at large times.
To do this we utilize large deviation theory \cite{T-09,MS-17}. We show that both distributions admit the large deviation form and compute corresponding rate functions. All analytical computations are supported by numerical simulations.

\clearpage

\par 
Let us stress that the analysis we are to conduct in this paper closely resembles the one presented in \cite{BMR-23} for the local time density at the origin. Moreover,  since $\delta(x) = \pd_x \theta(x)$, the occupation time and the local time are related in a simple way on the functional level. 
However, this similarity is, in fact, misleading, and one cannot derive the statistical properties of the occupation time from the known results for the local time density. Therefore to describe statistics of the occupation time precisely, one has to go through the whole machinery of computations. In the present paper we provide these calculations.

\par  
The paper is organized as follows. 
In Section~\ref{seq:model and results} we introduce the model and the problem we address and present our main results. 
In Section~\ref{sec:Mean and Variance}, we consider the case of general steplike initial condition. By computing first two cumulants of both quenched and annealed probability distributions, we describe the behavior of the occupation time close to its typical value. 
We show that the variance depends on the initial condition, and this dependence persists over time. We also show that this memory effect is governed by a single static quantity, known as the generalized compressibility or Fano factor.
In Section~\ref{seq:Large deviation functions} we focus on uncorrelated uniform initial conditions and characterize the tails of the probability distributions by computing corresponding large deviation functions along with their asymptotic expansions. 
Section~\ref{sec:numerics} is devoted to numerical simulations. 
Finally, we conclude in Section~\ref{sec:conclusion}. 
In addition, in Appendix~\ref{sec:app-OTvsLTD} we provide detailed comparison between the results presented in this paper and those obtained in \cite{BMR-23} for the local time density at the origin.

\section{The model and the main results}\label{seq:model and results}
\subsection{The model}\label{seq:model}
\begin{figure}[h]
\includegraphics[width = \linewidth]{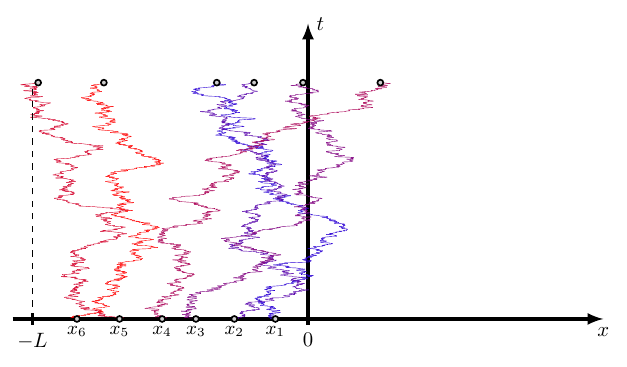}
\caption{Schematic representation of the Brownian trajectories of for $N=6$ particles.}
\label{fig:brownian}
\end{figure}
Consider a system of $N$ noninteracting Brownian particles on a line initially confined in the box $[-L;0]$ (see Fig~\ref{fig:brownian}). If we denote the coordinate of the $i$th particle by $x_i(t)$, then the evolution of the system is governed by $N$ independent Langevin equations
\begin{equation}\label{eq:evolution}
  \dot{x}_i(t) = \sqrt{2D} \, \eta_i(t), \qquad i=1,\ldots,N
\end{equation}
where $D$ is the diffusion coefficient and $\eta_i(t)$ is Gaussian white noise with zero mean and unit variance
\begin{equation}
  \left\langle \eta_i(t)\right\rangle = 0, \qquad
  \left\langle \eta_i(t) \eta_j(t') \right\rangle = \delta_{ij}\,\delta(t-t').
\end{equation}

\par  In this paper we study statistical properties of the occupation time (on the positive half-line). Denoting the total time of the process by $t$, we define this observable as 
\begin{equation}\label{eq:OT=def}
  T \equiv \sum_{i} \int_{0}^{t} \theta\left( x_i(t') \right) \dd t'
\end{equation}
where $\theta(x)$ is a Heaviside $\theta$-function. 
The quantity \eqref{eq:OT=def} measures the amount of time particles have spent to the right of the origin. 
Note, that in \cite{BMR-23} we used the same notation $T$ for the local time density at the origin, but these quantities do not appear together in the main text of the paper, therefore it should not cause any confusion.

\par Clearly, the occupation time depends on the initial positions of the particles. 
Therefore, in order for the problem to be well-defined, we need to specify them. In this paper we assume that initial coordinates of the particles are random variables which, in principle, may exhibit correlations. The only requirement is that the marginal probability distribution of a single particle initial coordinate is uniform along the interval $[-L,0]$. 

\par Since the initial positions of the particles are random, there are two sources of randomness in the system: stochasticity of Brownian trajectories and initial distribution of particles. 
The latter, as was argued in \cite{DG-09}, plays the role, similar to the realization of the disorder in the theory of the disordered systems. Therefore there are two natural ways to treat it. One can either average over all possible realizations  of the initial configurations (``\textit{annealed}'' averaging scheme), or consider a typical initial condition (``\textit{quenched}'' averaging scheme). Let us now endow these two schemes with formal definitions.

\par Consider some fixed initial configuration of particles $\mathbf{x}=(x_1,\ldots,x_N)$. Denote by $\mathbb{P}[T,t \, \vert\, \mathbf{x}]$ the probability that at the observation time $t$ the total occupation time is equal to $T$ and by $\left\langle e^{-pT} \right\rangle_\mathbf{x}$ its Laplace transform
\begin{equation}\label{eq:<e^-pT>=def N particles}
  \left\langle e^{-pT}\right\rangle_\mathbf{x} \equiv 
  \int_{0^{-}}^{Nt} \dd T\, e^{-pT} \mathbb{P}[T,t\,\vert\, \mathbf{x}].
\end{equation}
Then annealed and quenched probability distributions are defined as
\begin{align}
  \label{eq:P_an[T,t]=def}
  & \int_{0^{-}}^{Nt} \dd T\, e^{-pT} \mathbb{P}_\text{an}[T,t] \equiv 
  \overline{ \left\langle e^{-pT}  \right\rangle_\mathbf{x}},\\
  \label{eq:P_qu[T,t]=def}
  & \int_{0^{-}}^{Nt} \dd T\, e^{-pT} \mathbb{P}_\text{qu}[T,t] \equiv 
  \exp\left[ \overline{ \log \left\langle e^{-pT}  \right\rangle_\mathbf{x}}\right]
\end{align}
Here and subsequently bar $\overline{\cdots}$ denotes averaging over realizations of the initial conditions and brackets $\left\langle \cdots\right\rangle_\mathbf{x}$ stand for averaging over Brownian trajectories with initial configuration $\mathbf{x}$ kept fixed. The upper limit in the Laplace transforms is due to the fact that the occupation time for a single particle is bounded by the total observation time.

\par The aim of this paper is to study the statistics of the occupation time \eqref{eq:OT=def} at large times in the thermodynamic limit, i.e. the limit in which $N,L\to\infty$ with their ratio $\bar{\rho}=N/L$ being fixed. In particular we shall study two probability distributions of the occupation time: annealed $\mathbb{P}_\text{an}[T,t]$ and quenched $\mathbb{P}_\text{qu}[T,t]$, defined by  \eqref{eq:P_an[T,t]=def} and  \eqref{eq:P_qu[T,t]=def} respectively.

\subsection{The main results}\label{seq:results}
The probability distribution of the occupation time close to its typical value has a Gaussian form which can be characterized by the mean and the variance. We compute them for the general steplike initial condition in which initial coordinates may be correlated.

\par First, we compute the mean value of the occupation time 
\begin{equation}\label{eq:<T>=MainResults}
  \overline{ \left\langle T \right\rangle_\mathbf{x}  } =  \frac{2}{3}\sqrt{\frac{D}{\pi}} 
    \bar{\rho} \, t^{3/2}.
\end{equation}
Expression \eqref{eq:<T>=MainResults} holds for both annealed and quenched probability distributions. However the variances differ. For the quenched variance, which is defined in accordance with \eqref{eq:P_qu[T,t]=def} as
\begin{equation}\label{eq:Var_qu[T]=def MainResults}
  \mathrm{Var}_\text{qu}[T] =
    \overline{ 
      \left\langle T^2 \right\rangle_\mathbf{x} 
      - \left\langle T\right\rangle_\mathbf{x}^2
    },
\end{equation}
we find the explicit form 
\begin{equation}\label{eq:Var_qu[T]=answer MainResults}
  \mathrm{Var}_\text{qu}[T]
  = \frac{8 \left(\sqrt{2}-1\right) }{15} \sqrt{\frac{D}{\pi}}\, \bar{\rho} \, t^{5/2}.
\end{equation}
We emphasize that results \eqref{eq:<T>=MainResults} and \eqref{eq:Var_qu[T]=answer MainResults} hold for any time~$t$. Another observation we can make from \eqref{eq:Var_qu[T]=answer MainResults} is that the quenched variance depends only on the average density of particles in the initial configuration and has no memory of possible correlations whatsoever. 

The annealed variance according to \eqref{eq:P_an[T,t]=def} is defined as
\begin{equation}\label{eq:Var_an=def MainResults}
\mathrm{Var}_\text{an}[T] 
  =
  \overline{ \left\langle T^2 \right\rangle_\mathbf{x} } 
  - {\overline{ \left\langle T\right\rangle }_\mathbf{x}}^2.
\end{equation}
We find that at large times its behavior reads
\begin{equation}\label{eq:Var_an[T]=MainResults}
  \mathrm{Var}_\text{an}[T] 
  \underset{t\to\infty}{\simeq}
  \left[ \frac{2}{5} + \frac{2\,(4\sqrt{2}-7)}{15}(1-\alpha_\text{ic}) \right] \sqrt{\frac{D}{\pi}} \, \bar{\rho} \, t^{5/2},
\end{equation}
where $\alpha_\text{ic}$ denotes the Fano factor (generalized compressibility) of the initial condition
\begin{equation}\label{eq:alpha_ic=def MainResults}
  \alpha_\text{ic} \equiv \lim_{\ell\to\infty} \frac{ \mathrm{Var}[n(\ell)] }{ \overline{n(\ell)} }.
\end{equation}
Here $n(\ell)$ stands for the number of particles initially located in the segment $[-\ell, 0]$. 

\par It is clear from \eqref{eq:Var_an[T]=MainResults}, that the annealed variance does indeed depend on the initialization. However, all the dependence is encoded in a single static quantity $\alpha_\text{ic}$.
In other words, if we initialize the systems in two different ways --- by drawing the initial configurations from either distribution $p_1(\mathbf{x})$ or $p_2(\mathbf{x})$ with different Fano factor --- then we can distinguish between these initialization protocols by looking at the annealed variance of the occupation time. 
This is exactly what we mean by the memory of the initial condition.


{
\par Let us take a closer look at two particular cases. First is the hyperuniform initial condition, i.e. the initial condition in which the ratio $\ell^{-1}\mathrm{Var}[n(\ell)]\to 0$  as $\ell\to0$.  In this case $\alpha_\text{ic}$ is equal to zero. Consequently 
\begin{equation}\label{eq:Var_an[T] = (alpha=0)}
  \alpha_\text{ic} = 0 : \quad
     \mathrm{Var}_\text{an}[T] 
      \underset{t\to\infty}{\simeq}
      \frac{8 \left(\sqrt{2}-1\right) }{15} \sqrt{\frac{D}{\pi}}\, \bar{\rho} \, t^{5/2}.
\end{equation}
Comparing \eqref{eq:Var_an[T] = (alpha=0)} with \eqref{eq:Var_qu[T]=answer MainResults} suggests, that the ``typical'' initial configuration, as mentioned when handwavingly defining the quenched probability distribution, is essentially a configuration drawn from a hyperuniform distribution (see Sec~\ref{sec:Mean and Variance: N particles} for more details).     
}

\par Another important special case is uncorrelated uniform initial condition, the scenario in which initial positions of particles are drawn from the uniform distribution independently.  
Since particles do not interact, uncorrelated uniform distribution is essentially the equilibrium distribution and therefore it is a very natural choice for the initial condition. 
In this case, the distribution of the number of particles in a segment $n(\ell)$ is a Poisson distribution. Therefore the mean and the variance entering \eqref{eq:alpha_ic=def MainResults} are the same, hence $\alpha_\text{ic}=1$ and we have
\begin{equation}\label{eq:Var_an[T] = (alpha=1)}
\alpha_\text{ic} = 1: \quad
  \mathrm{Var}_\text{an}[T] 
  \underset{t\to\infty}{\simeq}
  \frac{2}{5} \sqrt{\frac{D}{\pi}} \, \bar{\rho} \, t^{5/2}.
\end{equation}
Actually, for the uncorrelated uniform initial condition, expression \eqref{eq:Var_an[T] = (alpha=1)} is exact for all times and not only in the limit $t
\to\infty$ (see \ref{sec:Mean and Variance: N particles} for details). In particular this implies, that 
\begin{equation}
  \text{uncorrelated uniform:}\qquad
  \frac{ \mathrm{Var}_\text{qu}[T]  }{\mathrm{Var}_\text{an}[T] } = 
  4 \frac{\sqrt{2} - 1}{3} < 1
\end{equation}
and this ratio \emph{ does not depend on time}.

\par To explore the tails of the probability distributions we resort to the large deviation formalism \cite{T-09,MS-17}. We find that the probability distributions admit the large deviation forms
\begin{align}
  \label{eq:P_an[T,t]=LD form MainResults}
  \mathbb{P}_\text{an}[T,t] & \simeq 
      \exp\left[ -\bar{\rho} \sqrt{4Dt} \; \Phi_\text{an}(\tau) \right], \\
  \label{eq:P_qu[T,t]=LD form MainResults}
  \mathbb{P}_\text{qu}[T,t] & \simeq
      \exp\left[ -\bar{\rho} \sqrt{4Dt} \; \Phi_\text{qu}(\tau) \right],
\end{align}
where  
\begin{equation}\label{eq:tau=def}
  \tau \equiv \frac{T}{t^{3/2}} \frac{1}{\bar{\rho}\sqrt{4D}}.
\end{equation}
As we show in Sec.~\ref{sec:Mean and Variance: N particles}, quenched rate function $\Phi_\text{qu}(\tau)$ is the same for all steplike initial conditions. Annealed rate function $\Phi_\text{an}(\tau)$, on the other hand, is more subtle, and is different for different initial conditions. Therefore we restrict ourselves to the case of uncorrelated uniform initial condition. 

\begin{figure*}
  \includegraphics[width = \linewidth]{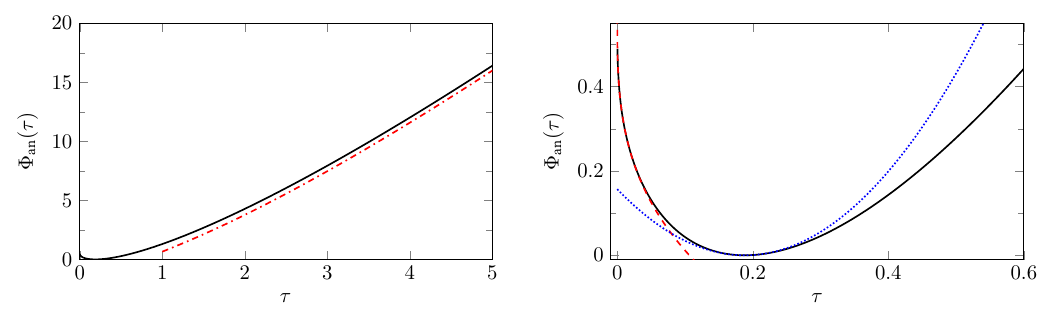}
 \caption{Large deviation function in the annealed case and its asymptotic expansions. In the left panel $\tau$ has a larger range, while in the right panel, we zoom in around the minimum of $\Phi_\text{an}(\tau)$ reached at $\tau=\tau_\text{typ} = \frac{1}{3\sqrt{\pi}}$. On both panels the solid lines correspond to the large deviation function computed in Mathematica from \eqref{eq:Phi_an(tau)=max() MainResults} and \eqref{eq:phi_an(tau)= MainResults}. 
 Dashed, dotted and dash-dotted lines are asymptotic behaviors of $\Phi_\text{an}(\tau)$ in \eqref{eq:Phi_an(tau)=asympt MainResults} for small, typical and large values of rescaled occupation time $\tau$ \eqref{eq:tau=def} respectively.}
 \label{fig:Phi_annealed_anal}
\end{figure*} 

\par For the uncorrelated uniform initial condition we find that the annealed rate function $\Phi_\text{an}(\tau)$ is given by an inverse Legendre transform 
\begin{equation}\label{eq:Phi_an(tau)=max() MainResults}
  \Phi_\text{an}(\tau) = \max_{q}\left( - q \tau + \phi_\text{an}(q) \right)
\end{equation}
with 
\begin{equation}\label{eq:phi_an(tau)= MainResults}
  \phi_\text{an}(q) = \frac{1}{\sqrt{\pi}} 
    - \frac{1}{\sqrt{4q}} \erf\left(\sqrt{q}\right).
\end{equation}
Expression \eqref{eq:phi_an(tau)= MainResults} together with \eqref{eq:Phi_an(tau)=max() MainResults} provide parametric representation of $\Phi_\text{an}(\tau)$ (see Fig~\ref{fig:Phi_annealed_anal}). Analyzing these two expressions we extract leading asymptotic behaviors of $\Phi_\text{an}(\tau)$
\begin{equation}\label{eq:Phi_an(tau)=asympt MainResults}
  \Phi_\text{an}(\tau) \sim
  \begin{cases}
     \frac{1}{\sqrt{\pi}} - \frac{3}{2}\left(\frac{\tau}{2}\right)^{1/3},
    {}& \tau \to 0,\\
    \frac{5}{2} \sqrt{\pi} \left(\tau - \frac{1}{3\sqrt{\pi}}\right)^{2},
    {}& \tau \to \frac{1}{3\sqrt{\pi}},
    \\
    \tau\left( \mu \log \left[ 2\tau \sqrt{\pi}\right]  
           - \frac{1}{\mu} \right),
    {}& \tau \to \infty,
  \end{cases}
\end{equation}
where 
\begin{equation}\label{eq:mu(tau)=MainResults}
  \mu \equiv 1 + \frac{ \log\log \left[ 2\tau \sqrt{\pi} \right]}
                          { \log \left[2\tau\sqrt{\pi}\right] }.
\end{equation}
Asymptotic expansion \eqref{eq:Phi_an(tau)=asympt MainResults} along with the large deviation form \eqref{eq:P_an[T,t]=LD form MainResults} imply that close to the typical value of the occupation time, annealed probability distribution is given by
\begin{multline}
  \mathbb{P}_\text{an}[T,t] 
    \simeq
  \exp \left[ 
    - \frac{1}{2} 
    \left( \frac{T - T_\text{typ}}{\sigma_\text{an}} \right)^2
  \right], \quad 
  T\to T_\text{typ} \\
\end{multline}
with 
\begin{equation}
  T_\text{typ} = \frac{2}{3} \,\sqrt{\frac{D}{\pi}}  \bar{\rho}\, t^{3/2},
  \qquad
  \sigma_\text{an}^2 = 
  \frac{2}{5}\, \sqrt{\frac{D}{\pi}}\, \bar{\rho}\,t^{5/2}.
\end{equation}
This is exactly the result we anticipate from \eqref{eq:<T>=MainResults} and \eqref{eq:Var_an[T] = (alpha=1)}.

\begin{figure*}
  \includegraphics[width = \linewidth]{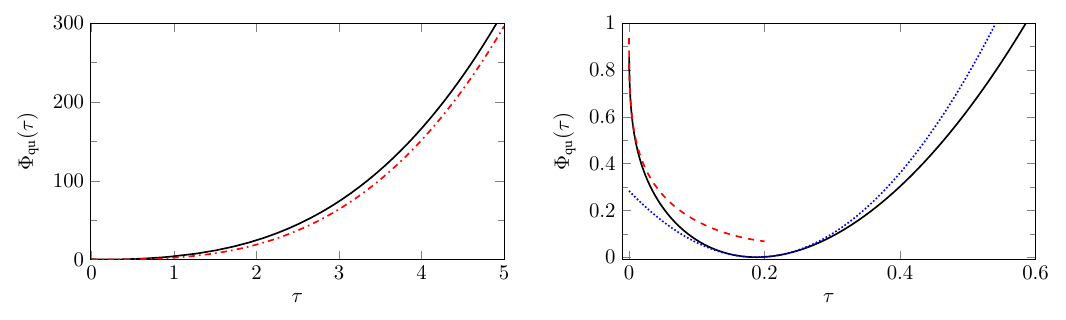}
 \caption{Large deviation function in the quenched case and its asymptotic expansions. In the left panel $\tau$ has a larger range, while in the right panel, we zoom in around the minimum of $\Phi_\text{qu}(\tau)$ reached at $\tau=\tau_\text{typ} = \frac{1}{3\sqrt{\pi}}$. On both panels the solid lines correspond to the large deviation function computed in Mathematica from \eqref{eq:Phi_qu(tau)=max() MainResults} and \eqref{eq:phi_qu(tau)=MainResults}. 
 Dashed, dotted and dash-dotted lines are asymptotic behaviors of $\Phi_\text{qu}(\tau)$ in \eqref{eq:Phi_qu(tau)=asympt MainResults} for small, typical and large values of rescaled occupation time $\tau$ \eqref{eq:tau=def} respectively.}
 \label{fig:Phi_quenched_anal}
\end{figure*} 

\par For the quenched rate function $\Phi_\text{qu}(\tau)$ we obtain similar representation. Namely
\begin{equation}\label{eq:Phi_qu(tau)=max() MainResults}
  \Phi_\text{qu}(\tau) = \max_q\left( - q \tau + \phi_\text{qu}(q)  \right)
\end{equation}
where 
\begin{equation}\label{eq:phi_qu(tau)=MainResults}
    \phi_\text{qu}(q) = 
    -\int_{0}^{\infty} \dd y
     \log \left[
      \erf(y) 
      + 
      \frac{1}{\pi} \int_{0}^{1} \dd \xi \frac{ e^{- q \xi - \frac{y^2}{1-\xi}} }{\sqrt{\xi(1-\xi)}} 
     \right].  
\end{equation}
Expressions \eqref{eq:Phi_qu(tau)=max() MainResults} and \eqref{eq:phi_qu(tau)=MainResults} give us parametric representation of the quenched rate function $\Phi_\text{qu}(\tau)$ (see Fig~\ref{fig:Phi_quenched_anal}) and allows us to find the asymptotic behaviors
\begin{equation}\label{eq:Phi_qu(tau)=asympt MainResults}
  \Phi_\text{qu}(\tau) \sim
  \begin{cases}
    \phi_{\infty} - 
    \frac{ 1 + \frac{1}{2} \log \nu  + \frac{1}{3}\log\log \nu}
         { \frac{2}{\sqrt{\pi}}\left(\nu \log \nu\right)^{1/3} },
    {}& \tau \to 0,\\
    \frac{1}{2}\frac{15\sqrt{\pi}}{4(\sqrt{2}-1)} \left(\tau - \frac{1}{3\sqrt{\pi}}\right)^{2},
    {}& \tau \to \frac{1}{3\sqrt{\pi}},
    \\
    \left(\frac{4}{3}\tau\right)^{3} ,
    {}& \tau \to \infty,
  \end{cases}
\end{equation}
where
\begin{equation}\label{eq:nu=, phi_infty= }
  \nu = \frac{\pi^{3/2}}{12\tau},
  \quad
  \phi_\infty = - \int_{0}^{\infty} \log\left[ \erf(y) \right] \dd y \approx 1{.}03.
\end{equation}
Similarly to the annealed case, from \eqref{eq:Phi_qu(tau)=asympt MainResults} and \eqref{eq:P_qu[T,t]=LD form MainResults} it is clear that close to the typical value of the occupation time  
\begin{equation}
  \mathbb{P}_\text{qu}[T,t] \simeq \exp\left[ - \frac{1}{2} \left( \frac{T - T_\text{typ}}{\sigma_\text{qu}}\right)^2 \right], \quad 
  T \to T_\text{typ}
\end{equation}
where
\begin{equation}
  T_\text{typ} = \frac{2}{3} \,\sqrt{\frac{D}{\pi}}  \bar{\rho}\, t^{3/2},
  \quad
  \sigma_\text{qu}^2 = \frac{8 \left(\sqrt{2}-1\right) }{15} \sqrt{\frac{D}{\pi}}\, \bar{\rho} \, t^{5/2}. 
\end{equation}
which indeed matches with the mean \eqref{eq:<T>=MainResults} and the variance \eqref{eq:Var_qu[T]=answer MainResults}.

\begin{figure}
  \includegraphics[width = \linewidth]{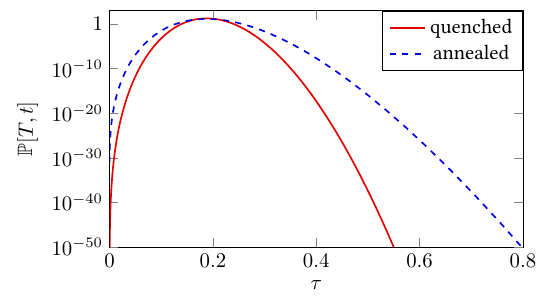}
  \caption{Quenched (solid) and annealed (dashed) probability distributions obtained from \eqref{eq:P_qu[T,t]=LD form MainResults} and \eqref{eq:P_an[T,t]=LD form MainResults} respectively. In this plot we use $t=10000$, $D=1/2$, $\bar{\rho}=1$.}
  \label{fig:quenchedVSannelaed}
\end{figure}
\par Note that the annealed and quenched probability distributions are very different (see Fig~\ref{fig:quenchedVSannelaed} for the comparison). For example, the probabilities that $T=0$, i.e. the probabilities that no particle have reached the origin up to the observation time $t$, in both cases are stretched exponentials
\begin{equation}\label{eq:P_an[0,t]= P_qu[0,t]= str exp}
  \mathbb{P}_\text{an}[0,t] \sim e^{ - \theta_\text{an} \, \bar{\rho} \sqrt{4Dt}  },
  \qquad
  \mathbb{P}_\text{qu}[0,t] \sim e^{ - \theta_\text{qu} \, \bar{\rho} \sqrt{4Dt}  },
\end{equation}
but the constants are different. Namely 
\begin{equation}\label{eq:theta_an= theta_qu=}
   \theta_\text{an} = \frac{1}{\sqrt{\pi}}\approx 0{.}56,\qquad
   \theta_\text{qu} = \phi_\infty \approx 1{.}03.
\end{equation} 
This difference is due to the atypical initial conditions. The annealed probability distribution takes into account initial configurations in which particles are far from the origin and hence they require more time to reach it, which leads to higher probability of survival. Such initial conditions are indeed the atypical ones and therefore they do not contribute to the quenched probability distribution. 

\par In fact, survival probabilities \eqref{eq:P_an[0,t]= P_qu[0,t]= str exp} naturally appear in the context of target problems (see \cite{BMS-13} for a review) and both constants in \eqref{eq:theta_an= theta_qu=} can be computed in a much simpler way \cite{MVP-14}. Here we obtain them as byproducts.

\par The behaviors of the quenched and annealed distributions at atypically large occupation times also differ. Asymptotic expansions \eqref{eq:Phi_an(tau)=asympt MainResults} and \eqref{eq:Phi_qu(tau)=asympt MainResults} imply that
\begin{align}
    & \mathbb{P}_\text{an}[T,t] 
     \sim 
    \exp\left[ - {\textstyle \frac{1}{2} \frac{T}{t}} 
      \log\left( {\textstyle\frac{1}{t^3D\bar{\rho}^2}}\, T^2 \right) \right],
     &&\! T \gg T_\text{typ},
    \\
    & \mathbb{P}_\text{qu}[T,t] \sim 
    \exp\left[ - {\textstyle \frac{16}{27}\frac{T}{t}} 
      \left( {\textstyle\frac{1}{t^3D\bar{\rho}^2}}\, T^2 \right) \right],
    &&\!
     T \gg T_\text{typ}.
\end{align}  
Thus, the quenched probability distribution decay much faster than the annealed one. This is due to the atypical initial configurations, in which particles are concentrated in the vicinity of the origin. Since it does not take much time for these particles to reach the origin, they contribute to the atypically large values of the occupation time.

\section{The mean and the variance}\label{sec:Mean and Variance}
\par Close to the typical value, we expect the probability distribution of the occupation time to be Gaussian and hence we can describe it by the mean and the variance. In this section we compute them for general steplike initial condition in both annealed and quenched averaging schemes. 

\subsection{One particle}\label{sec:Mean and Variance: 1 particle}
\par Let us first consider the case of the individual particle. In the case of simple Brownian motion the probability distribution $\mathbb{P}[T,t\,\vert\,x]$ of the occupation time $T$ can be found e.g. in \cite{Borodin-Handbook} (p. 162 therein), and for a pedagogical derivation of this result we refer the reader to \cite{M-05-BFPCS}. If the particle starts at $x(0)=x\le0$, then the probability distribution  $\mathbb{P}[T,t\,\vert\,x]$ reads
\begin{multline}\label{eq:P[T,t|x]=1p}
  \mathbb{P}[T,t \,\vert\, x] = 
    \erf\left[ \frac{\left| x \right|}{\sqrt{4Dt}} \right]\delta(T)
    \\
    +
    \frac{1}{\pi} \frac{1}{\sqrt{T(t-T)}} \exp \left[ -\frac{x^2}{4D(t-T)} \right],
    \quad x \le 0,
\end{multline} 
where
\begin{equation}
  \erf(z) = \frac{2}{\sqrt{\pi}}\int_{0}^z e^{-s^2} \dd s.
\end{equation}
Note, that if the particle starts at the origin, then \eqref{eq:P[T,t|x]=1p} simplifies into the famous L\'evy's arcsine law \cite{L-40}
\begin{equation}
  \mathbb{P}[T,t\,\vert\,x=0] = \frac{1}{\pi} \frac{1}{\sqrt{T(t-T)}}.
\end{equation}
The first term in \eqref{eq:P[T,t|x]=1p} comes from the trajectories that have not reached the origin up to time $t$, and is nothing but the survival probability of an individual particle. 

\par We should stress, that the expression \eqref{eq:P[T,t|x]=1p} is valid only when $x\le0$. But since this is exactly the case we are interested in, to simplify the notation we shall not emphasize this anymore. 

\par We will also need the Laplace transform of the $\mathbb{P}[T,t \,\vert\, x]$ which is defined in a usual way as
\begin{equation}\label{eq:<e^-pT>=def}
  \left\langle e^{-pT} \right\rangle_x \equiv 
  \int_{0^-}^{t} e^{-pT}\, \mathbb{P}[T,t \,\vert\, x]\, \dd T.
\end{equation}
Since the occupation time cannot be larger than $t$, the probability measure is supported on $[0,t]$ and hence the upper limit in the integral above.
Using the explicit form \eqref{eq:P[T,t|x]=1p} of the probability distribution $\mathbb{P}[T,t\,\vert\,x]$ we obtain 
\begin{multline}\label{eq:<e^-pT>=int_form}
  \left\langle e^{-pT} \right\rangle_x 
  = \erf\left[ \frac{\left| x \right|}{\sqrt{4Dt}} \right]
  \\
  + 
  \frac{1}{\pi} \int_{0}^{1} \dd \xi \frac{ e^{- p t \, \xi} }{\sqrt{\xi(1-\xi)}} 
  \exp \left[ -\frac{x^2}{4Dt}\frac{1}{(1-\xi)} \right].
\end{multline}
The integral in \eqref{eq:<e^-pT>=int_form} has no explicit expression in terms of elementary functions, but nevertheless it provides us  a lot of information. For example the moments of the occupation time are given by
\begin{equation}\label{eq:<T^n>=def}
   \left\langle T^n \right\rangle_x = 
   t^n \, \frac{1}{\pi} 
   \int_{0}^1 \dd \xi \frac{\xi^n}{\sqrt{\xi(1-\xi)}} 
   \exp \left[ -\frac{x^2}{4Dt}\frac{1}{(1-\xi)} \right].
\end{equation} 
After the change of variables $\xi = \frac{y}{y+1}$ and rescaling of the coordinate $z = \frac{x^2}{4Dt}$ expression \eqref{eq:<T^n>=def} transforms into
\begin{equation}\label{eq:<T>=Psi(..,..,..)}
  \left\langle  T^{n} \right\rangle_{x=-\sqrt{4Dt\, z}} = t^n \, 
  e^{ - z }
  \frac{\Gamma\left(n+\frac{1}{2}\right) }{\pi}
  U\left(  
      n+\frac{1}{2}, \frac{1}{2}; z
  \right),
\end{equation}
where $\Gamma(a)$ is Euler gamma function and $U(a,b;z)$ denotes the confluent hypergeometric function of the second kind \cite{Higher-trans-funct-vol1}
\begin{equation}\label{eq:con_hg_function=def}
  U(a,b; z) \equiv \frac{1}{\Gamma(a)} \int_{0}^{\infty} \dd s\, 
    e^{-z s} s^{a-1} (s+1)^{b-a-1} 
\end{equation}
From \eqref{eq:con_hg_function=def} it follows that $U(a,b;z)$ satisfies a recurrence relation
\begin{equation}\label{eq:Psi_recurrence}
  U(a+1,b;z) = \frac{z^{1-a}}{a(a-b+1)} \, \dv{}{z}\Big[  z^a U(a,b; z)\Big].
\end{equation}
In addition, one can easily check that 
\begin{equation}\label{eq:Psi(1/2,1/2)}
  U\left(\frac{1}{2},\frac{1}{2}; z \right) = \sqrt{\pi} e^{ z } \left( 1 - \erf(\sqrt{z}) \right).
\end{equation}
Combining \eqref{eq:Psi(1/2,1/2)} with recurrence relation \eqref{eq:Psi_recurrence} and then using \eqref{eq:<T>=Psi(..,..,..)} we can find all the moments of the occupation time. In particular the first two moments are
\begin{align}
  \label{eq:<T>=1p}
  \left\langle T\right\rangle_{x=-\sqrt{4Dt\, z}}  {}&  = 
    t \left[ \left(\frac{1}{2} + z \right) \erfc (\sqrt{z}) - \frac{\sqrt{z} e^{-z}}{\sqrt{\pi}} \right],
  \\
  \nonumber
  \left\langle T^2\right\rangle_{x=-\sqrt{4Dt\, z}} {}&  = t^2
  \left(\frac{3}{8}+\frac{3}{2}\,z+\frac{1}{2}\,z^2\right) \erfc (\sqrt{z})
  \\ 
  \label{eq:<T2>=1p}
  & \qquad \qquad - 
       t^2 \left(\frac{1}{2}\,z + \frac{5}{4}\right) \frac{\sqrt{z} e^{-z}}{\sqrt{\pi}}
\end{align}
where  $\erfc(z) = 1 - \erf(z)$. From \eqref{eq:<T>=1p} and \eqref{eq:<T2>=1p} we extract large time behavior for the mean and the variance of the occupation time. At large times 
\begin{equation}
  z = \frac{x^2}{4Dt} \to 0, \qquad t \to \infty
\end{equation}
and hence
\begin{equation}\label{eq:1p mean and var scaling}
  \left\langle T\right\rangle_x  \sim \frac{1}{2}\, t, 
  \qquad
  \left\langle T^2\right\rangle_x - \left\langle T\right\rangle_x^2 \sim \frac{1}{8}\, t^2, 
  \qquad
  t \to \infty.
\end{equation}

\subsection{System of particles}\label{sec:Mean and Variance: N particles}
Now we proceed to the system of $N$ Brownian particles initially located to the left of the origin and compute the mean and the variance of the occupation time. But first of all we shall specify initial conditions more precisely.

\par Denoting by $x_i$ the position of the $i$-th particle at $t=0$, we define the empirical density of the initial condition as
\begin{equation}\label{eq:rho_emp = def}
  \hat{\rho}(y\,\vert\,\mathbf{x}) \equiv \sum_{i} \delta( x_i - y ).
\end{equation}
Now all the information about the initial configuration is encoded in $\hat{\rho}(y\,\vert\,\mathbf{x})$. In what follows, we restrict ourselves to the particular type of the initial condition, that is we assume that the initial configuration of the particles was obtained from the translational invariant distribution on the real line by removing all particles on the positive half-line. This assumption essentially implies two things. First, the average density is a constant to the left of the origin and zero to the right
\begin{equation}\label{eq:rho=ass1}
  \overline{\hat{\rho}(y\,\vert\,\mathbf{x})} = \bar{\rho} \; \theta(-y).
\end{equation}
Second, two point correlation function $C(y_1,y_2)$ of the initial condition
\begin{equation}\label{eq:C(y1,y2)=def}
  C(y_1,y_2) \equiv \overline{ \hat{\rho}(y_1\,\vert\,\mathbf{x}) \, \hat{\rho}(y_2\,\vert\,\mathbf{x}) } - {\bar{\rho} }^2
\end{equation}
depends only on the difference $(y_1 - y_2)$
\begin{equation}\label{eq:rho=ass2}
  C(y_1,y_2) = \bar{\rho}\; \theta(-y_1) \theta(-y_2) \, C(y_1 - y_2).
\end{equation}

\par Note that we can already guess the scaling behavior of both the mean and the variance. Indeed, the typical displacement of a Brownian particle is proportional to $\sqrt{Dt}$. Consequently, the particles that have reached the origin up to time $t$ are those that were initially located within the segment $[-\sqrt{Dt},0]$. Since there are approximately $\bar{\rho} \, \sqrt{Dt}$ such particles, and contribution of each particle to the mean value of the occupation time is proportional to $t$ \eqref{eq:1p mean and var scaling}, we expect the mean value of the total occupation time to scale as $\sqrt{t}\, \cdot\, t = t^{3/2}$. Similarly, we expect the variance to scale as $\sqrt{t}\,\cdot\, t^2 = t^{5/2}$. Having this heuristic argument in mind, we proceed to the rigorous computation.

The mean value of the occupation time for a given initial configuration $\mathbf{x}$ in terms of the empirical density reads
\begin{equation}\label{eq:<T>_x=int rho}
  \left\langle T\right\rangle_\mathbf{x} = 
  \int_{-\infty}^{\infty} \dd y\, \hat{\rho}(y \, \vert\, \mathbf{x}) \left\langle T\right\rangle_y,
\end{equation}
where $\left\langle T\right\rangle_y$ denotes the mean value of the occupation time of a single particle starting at $y$ which is given by \eqref{eq:<T>=1p}. Averaging \eqref{eq:<T>_x=int rho} over initial conditions we get
\begin{equation}\label{eq:<T>_x = int bar rho}
  \overline{ \left\langle T\right\rangle_\mathbf{x} } = 
  \int_{-\infty}^{\infty} \dd y\, \overline{\hat{\rho}(y\,\vert\,\mathbf{x})} \left\langle T\right\rangle_y.
\end{equation}
Combining \eqref{eq:<T>_x = int bar rho} with \eqref{eq:rho=ass1} we obtain
\begin{equation}\label{eq:<T>_x = bar rho int }
  \overline{ \left\langle T\right\rangle_\mathbf{x} } = 
  \bar{\rho} \int_{-\infty}^{0} \dd y\, \left\langle T\right\rangle_y.
\end{equation}
Recalling \eqref{eq:<T>=1p} and after straightforward computation we arrive at
\begin{equation}\label{eq:<T>=t^3/2 answer}
  \overline{ \left\langle T\right\rangle_\mathbf{x} } = 
  \frac{2}{3}\sqrt{\frac{D}{\pi}}\, \bar{\rho}\, t^{3/2}.
\end{equation}
This is the result stated in \eqref{eq:<T>=MainResults}.
The mean value depends only on the average density of the particles and is the same for both quenched and annealed distributions. Let us now proceed to the computation of the variances

We start with the quenched distribution. According to~\eqref{eq:P_qu[T,t]=def}, the quenched variance is given by
\begin{equation}\label{eq:Var_qu=def}
  \mathrm{Var}_\text{qu}[T] =
  \overline{ \left\langle T^2\right\rangle_\mathbf{x} - \left\langle T\right\rangle_\mathbf{x}^2 } 
\end{equation}
or in terms of the empirical density
\begin{equation}
  \mathrm{Var}_\text{qu}[T] = \int_{-\infty}^\infty \dd y\, 
  \overline{ \hat{\rho}(y\,\vert\,\mathbf{x}) }
    \left[
      \left\langle T^2\right\rangle_y - 
      \left\langle T\right\rangle_y^2
    \right].
\end{equation}
Due to \eqref{eq:rho=ass1}, averaging over initial conditions yields
\begin{equation}\label{eq:Var_qu=rho int}
  \mathrm{Var}_\text{qu}[T] = 
  \bar{\rho}
  \int_{-\infty}^0 \dd y\, 
    \left[
      \left\langle T^2\right\rangle_y - 
      \left\langle T\right\rangle_y^2
    \right].
\end{equation}
Substituting \eqref{eq:<T>=1p} and \eqref{eq:<T2>=1p} into \eqref{eq:Var_qu=rho int} we get
\begin{equation}\label{eq:Var_qu=ans t^5/2}
  \mathrm{Var}_\text{qu}[T] = \frac{8 \left(\sqrt{2}-1\right) }{15} \sqrt{\frac{D}{\pi}}\, \bar{\rho} \, t^{5/2}.
\end{equation}
This is exactly the result \eqref{eq:Var_qu[T]=answer MainResults}.

Note that the quenched variance depends only on the average density of particles in the initial configuration. In fact, the same is true for the full probability distribution $\mathbb{P}_\text{qu}[T,t]$. It becomes clear if we rewrite its definition \eqref{eq:P_qu[T,t]=def} in terms of the empirical density \eqref{eq:rho_emp = def}. Since there is no interaction we have
\begin{equation}
  \log \left\langle e^{-pT}\right\rangle_{\mathbf{x}} = 
  \int_{-\infty}^{\infty} \dd y\, \hat{\rho}(y\,\vert\,\mathbf{x})\log \left\langle e^{-pT}\right\rangle_{y}
\end{equation}
hence
\begin{equation}
  \overline{ \log \left\langle e^{-pT}\right\rangle_{\mathbf{x}} } = 
  \bar{\rho} \int_{-\infty}^{0} \dd y\, \log \left\langle e^{-pT}\right\rangle_{y}
\end{equation}
and therefore
\begin{equation}\label{eq:P_qu=exp[rho]}
  \int_{0^{-}}^{Nt} \dd T\, e^{-pT} \mathbb{P}_\text{qu}[T,t] = 
  \exp\left[ \bar{\rho} \int_{-\infty}^{0} \dd y\, \log \left\langle e^{-pT}  \right\rangle_{y}\right].
\end{equation}

Now we advance to the annealed probability distribution. In accordance with \eqref{eq:P_an[T,t]=def} its variance is 
\begin{equation}\label{eq:Var_an=def}
  \mathrm{Var}_\text{an}[T] = 
  \overline{ \left\langle T^2\right\rangle_\mathbf{x} } 
  - 
  \overline{ \left\langle T\right\rangle_\mathbf{x} }^2.
\end{equation}
Following the procedure introduced in \cite{BJC-22} (see also \cite{DMS-23} Sec. IV) we formally rewrite \eqref{eq:Var_an=def} as a sum of two terms
\begin{multline}\label{eq:Var_an=Var_qu+Var_ic}
  \mathrm{Var}_\text{an}[T] 
  =
  \left[
  \overline{ \left\langle T^2\right\rangle_\mathbf{x} } 
  - \overline{\left\langle T\right\rangle_\mathbf{x}^2 }
  \right] 
  + 
  \left[ \overline{ \left\langle T\right\rangle_\mathbf{x}^2 }
  -\overline{ \left\langle T\right\rangle_\mathbf{x} }^2
  \right] \\
  = \mathrm{Var}_\text{qu}[T] + \mathrm{Var}_\text{ic}[T].
\end{multline}
The first term in \eqref{eq:Var_an=Var_qu+Var_ic} is the quenched variance \eqref{eq:Var_qu=def} and all dependence of possible correlations in the initial condition is encoded solely in the second term
\begin{equation}\label{eq:Var_ic=def}
  \mathrm{Var}_\text{ic}[T] \equiv 
  \overline{ \left\langle T\right\rangle_\mathbf{x}^2 }
  -\overline{ \left\langle T\right\rangle_\mathbf{x} }^2.
\end{equation}
To express $\mathrm{Var}_\text{ic}[T]$ in terms of of the empirical density \eqref{eq:rho_emp = def} we note that
\begin{equation}\label{eq:<T^2>=int int}
  \overline{ \left\langle T\right\rangle_\mathbf{x}^2 } = 
  \int_{-\infty}^{\infty} \dd y\, 
  \int_{-\infty}^{\infty} \dd y'\, 
  \overline{ \hat{\rho}(y\,\vert\,x) \, \hat{\rho}(y'\,\vert\,x)  }
  \left\langle T\right\rangle_{y} \left\langle T\right\rangle_{y'}
\end{equation}
and
\begin{equation}\label{eq:<T>^2 = int * int}
  \overline{ \left\langle T\right\rangle_\mathbf{x}}^2 = 
  \bar{\rho}^2
  \int_{-\infty}^{0} \dd y\, 
  \left\langle T\right\rangle_{y} \;
  \int_{-\infty}^{0} \dd y'\, 
  \left\langle T\right\rangle_{y'}.
\end{equation}
Substituting \eqref{eq:<T>^2 = int * int} and \eqref{eq:<T^2>=int int} into \eqref{eq:Var_ic=def} we express $\mathrm{Var}_\text{ic}[T]$ in terms of two point correlation function \eqref{eq:C(y1,y2)=def} as
\begin{equation}\label{eq:Var_ic=intintC(y)}
  \mathrm{Var}_\text{ic}[T] = \bar{\rho} 
  \int_{-\infty}^{0} \dd y \int_{-\infty}^{0} \dd y'
  \left\langle T\right\rangle_{y} \left\langle T\right\rangle_{y'} C(y-y').
\end{equation}
To proceed further we should either specify initial condition and provide the exact form of the $C(y-y')$ or, alternatively, we can study the large time limit and compute \eqref{eq:Var_ic=intintC(y)} as $t\to\infty$. The latter can be done by introducing the Fourier transform of the two-point correlation function
\begin{equation}\label{eq:C(y)=Fourier}
   C(y) = \frac{1}{2\pi} \int_{-\infty}^{\infty} \dd q\, e^{ \ii q y } S(q).
\end{equation}
Then \eqref{eq:Var_ic=intintC(y)} turns into
\begin{multline}\label{eq:Var_ic=int int int}
  \mathrm{Var}_\text{ic}[T] = \bar{\rho} 
    \int_{-\infty}^{0} \dd y\, \int_{-\infty}^{0} \dd y' 
  \left\langle T\right\rangle_y \left\langle T\right\rangle_{y'} 
  \\
  \frac{1}{2\pi}\int_{-\infty}^{\infty} \dd q\,
   S(q) e^{\ii q (y-y')}.
\end{multline}
In the limit $t\to\infty$ the integral over $q$ is essentially a $\delta$-function. Indeed, if we rescale the variables of the integration in \eqref{eq:Var_ic=int int int} as $y= 4Dt\, \tilde{y}$, $y' = 4Dt\, \tilde{y}'$, $q=\frac{p}{4Dt}$, then the integral over $q$ transforms into
\begin{equation}\label{eq:int dq S(q) sim delta}
  \frac{1}{2\pi}\int_{-\infty}^{\infty} \dd p\,
     S\left(\frac{p}{4Dt}\right) e^{\ii p (\tilde{y}-\tilde{y}')}
  \underset{t\to\infty}{\simeq}  S(0)\; \delta(\tilde{y}-\tilde{y}')
\end{equation}
where we used a representation 
\begin{equation}\label{eq:delta(z-z')=int}
  \delta(\tilde{y}-\tilde{y}') = 
  \frac{1}{2\pi}\int_{-\infty}^{\infty} \dd p\,
     e^{\ii p (\tilde{y}-\tilde{y}')}.
\end{equation}
Due to \eqref{eq:int dq S(q) sim delta} at large times \eqref{eq:Var_ic=int int int} simplifies into
\begin{equation}
  \mathrm{Var}_\text{ic}[T] \simeq \bar{\rho}\, S(0) \int_{-\infty}^{0} \dd y\, \left\langle T\right\rangle_y^2,\quad
  t\to\infty
\end{equation}
and using \eqref{eq:<T>=1p} we arrive at
\begin{equation}\label{eq:Var_ic=S(0)exact}
  \mathrm{Var}_\text{ic}[T] \simeq S(0)\,
    \frac{2(7-4\sqrt{2})}{15}\sqrt{\frac{D}{\pi}} 
    \,\bar{\rho}\,t^{5/2}, \quad t\to\infty.
\end{equation}
The quantity $S(0)$ is exactly the Fano factor defined in \eqref{eq:alpha_ic=def MainResults}. To see this, we express $n(\ell)$  in terms of the empirical density 
\begin{equation}
  n(\ell) = \int_{-\ell}^{0} \dd y \, \hat{\rho}\left(y\,\vert\,\mathbf{x}\right).
\end{equation}
hence 
\begin{equation}
  \overline{n(\ell)} \sim \ell\bar{\rho},\quad \ell\to\infty.
\end{equation}
Similarly for the variance 
\begin{equation}
  \mathrm{Var}[n(\ell)] 
  =
  \int_{-\ell}^{0} \dd y \int_{-\ell}^{0}\dd y'\,
    \left[ \overline{   \hat{\rho}\left(y\,\vert\,\mathbf{x}\right) 
              \, \hat{\rho}\left(y'\,\vert\,\mathbf{x}\right) }
    - \bar{\rho}^2 \right]
\end{equation}
the integrand is nothing but a two-point correlation function \eqref{eq:C(y1,y2)=def}. Using the Fourier transform \eqref{eq:C(y)=Fourier} after rescaling $p = q\ell$, $y=\tilde{y}\ell$, $y'=\tilde{y}'\ell$ we get 
\begin{equation}
  \mathrm{Var}[n(\ell)] 
  =\bar{\rho}\,\ell 
  \int_{-1}^{0} \dd \tilde{y} \int_{-1}^{0}\dd \tilde{y}'
  \int_{-\infty}^{\infty}\frac{\dd p}{2\pi}
    e^{\ii p (\tilde{y}-\tilde{y}')} S\left(\frac{p}{\ell}\right)
\end{equation}
which in the limit $\ell\to\infty$ reduces to
\begin{equation}
  \mathrm{Var}[n(\ell)] 
  \sim
  \bar{\rho}\,\ell 
  \int_{-1}^{0} \dd \tilde{y} \int_{-1}^{0}\dd \tilde{y}'
  \delta(\tilde{y}-\tilde{y}')
  S\left(0\right)
\end{equation}
hence
\begin{equation}
  \mathrm{Var}[n(\ell)] \sim \bar{\rho}\,\ell\, S(0)\quad \ell \to \infty
\end{equation}
and
\begin{equation}
  \alpha_\text{ic} \equiv \lim_{\ell\to\infty} \frac{\mathrm{Var}[n(\ell)]}{\overline{n(\ell)}}  = S(0).
\end{equation}
Having expressions \eqref{eq:Var_qu=ans t^5/2} and \eqref{eq:Var_ic=S(0)exact} for both terms in \eqref{eq:Var_an=Var_qu+Var_ic} we combine them together arriving at the result stated in \eqref{eq:Var_an[T]=MainResults}, i.e.
\begin{equation}\label{eq:Var_an=answer}
  \mathrm{Var}_\text{an}[T] \underset{t\to\infty}{\simeq}
  \left[ \frac{2}{5} + \frac{2\,(4\sqrt{2}-7)}{15}\left(1-\alpha_\text{ic}\right) \right] \sqrt{\frac{D}{\pi}} \, \bar{\rho} \, t^{5/2}.
\end{equation}

As we have mentioned, instead of taking the limit $t\to\infty$ in \eqref{eq:Var_ic=intintC(y)} we can choose a particular initial condition. For example, if we draw initial coordinates independently from the uniform distribution, then two point correlation function is just a $\delta$-function
\begin{equation}\label{eq:C(y1,y2)=delta}
  C(y_1,y_2) = \bar{\rho}\; \theta(-y_1) \theta(-y_2) \, \delta(y_1 - y_2).
\end{equation}
Then $\alpha_\text{ic} = S(0) = 1$ and \eqref{eq:Var_ic=intintC(y)} simplifies into
\begin{equation}\label{eq:Var_ic=int uncorrelated}
    \text{uncorr. uniform:}\quad
    \mathrm{Var}_\text{ic}[T] = \bar{\rho} 
    \int_{-\infty}^{0} \dd y\,  {\left\langle T\right\rangle_y}^2 
\end{equation}
hence
\begin{equation}\label{eq:Var_ic=1 * exact}
  \text{uncorr. uniform:}\quad
  \mathrm{Var}_\text{ic}[T] = 
    \frac{2(7-4\sqrt{2})}{15}\sqrt{\frac{D}{\pi}} 
    \,\bar{\rho}\,t^{5/2}.
\end{equation}
and for the annealed variance we get
\begin{equation}\label{eq:Var_an=answer uncorrelated}
  \text{uncorr. uniform:}\quad
  \mathrm{Var}_\text{an}[T] =
  \frac{2}{5} \sqrt{\frac{D}{\pi}} \, \bar{\rho} \, t^{5/2}.
\end{equation}
We stress that for the uncorrelated uniform initial condition, expression \eqref{eq:Var_an=answer uncorrelated} is valid at any time whereas \eqref{eq:Var_an=answer} is the behavior at large times.

\begin{figure*}
\includegraphics[width=\linewidth]{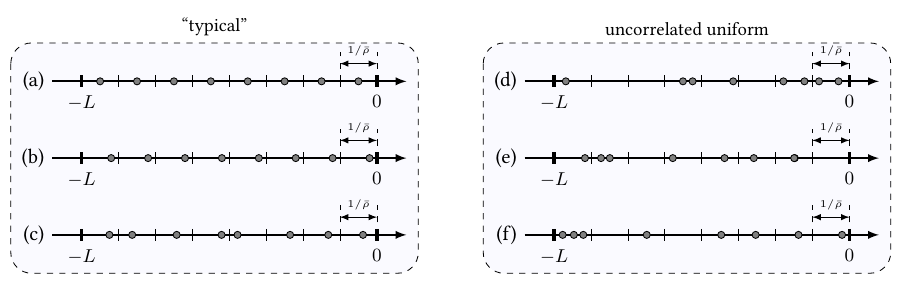}
\caption{Examples of typical initial configurations (a)-(c). For the comparison, we provide three initial configurations (d)-(f) drawn from the uncorrelated uniform probability distribution. }\label{fig:initial_conf}
\end{figure*}

{
\par As a final remark let us briefly comment on the physical meaning of the quenched probability distribution. At this point, we have already mentioned a couple of times that the quenched probability distribution corresponds to fixing the initial configuration of particles to be a ``typical'' one. But what are these ``typical'' configurations exactly? Comparing \eqref{eq:Var_an=answer} with \eqref{eq:Var_qu=ans t^5/2} suggests, that ``typical'' configurations are those, drawn from the probability distribution with $\alpha_\text{ic} = 0$. 
In fact, we can point out the ``typical'' initial conditions more precisely. 
Namely, if we consider a fixed initial configuration, such that $x_i(0)$ is confined within the segment $\chi_i$
\begin{equation}\label{eq:chi_segment=def}
  \chi_i \equiv \left[ - \frac{i}{\bar{\rho}}, - \frac{i-1}{\bar{\rho}} \right],   
\end{equation}
and study the probability distribution of the occupation time, then at large observation time, the resulting distribution will be the same as the quenched one $\mathbb{P}_{\text{qu}}\left[T,t\right]$ defined in \eqref{eq:P_qu[T,t]=def}. In other words all these configurations are ``typical'' (see Fig.~\ref{fig:initial_conf}). 
The rigorous proof of this statement is essentially the same as the one we have performed in \cite{BMR-23} for the local time density at the origin (see Appendix B therein), therefore, to avoid burying the reader under technical details, we do not provide the derivation here. 

\par A simple consequence is that if we draw $x_i(0)$ from a distribution supported on $\chi_i$, then all initial configurations will be ``typical'', and hence, at large times, they all result in $\mathbb{P}_{\text{qu}}\left[T,t\right]$. Consequently, within this initialization protocol, the quenched and the annealed probability distributions are the same. 
}

\section{The large deviation functions}\label{seq:Large deviation functions}
\par In the previous section we have described the behavior of the occupation time close to its typical value in both annealed and quenched cases for general steplike initial conditions. In this section we focus on the uncorrelated uniform initial condition and study the tails of the probability distributions \eqref{eq:P_an[T,t]=def} and \eqref{eq:P_qu[T,t]=def}. 
\par The mean and the variances we have computed suggest that at large times  probability distributions admit the following forms
\begin{align}
  \label{P_an[T,t]=LD_form}
  & \mathbb{P}_\text{an}[T,t] \simeq \exp\left[ -\bar{\rho}\sqrt{4Dt}\; \Phi_\text{an}(\tau)\right], \\
  \label{P_qu[T,t]=LD_form}
  & \mathbb{P}_\text{qu}[T,t] \simeq \exp\left[ -\bar{\rho}\sqrt{4Dt}\; \Phi_\text{qu}(\tau)\right],
\end{align}
with 
\begin{equation}
  \tau \equiv \frac{T}{t^{3/2}} \frac{1}{\bar{\rho}\sqrt{4D}}.
\end{equation}
In what follows we compute rate functions $\Phi_\text{an}(\tau)$ and $\Phi_\text{qu}(\tau)$ along with their asymptotic expansions.

\subsection{Annealed large deviation function}
As we have shown in \eqref{eq:Var_an=answer} annealed probability distribution depends on the initial condition, and this dependence does not vanish with time. Clearly such dependence remains on the level of the rate function $\Phi_\text{an}(\tau)$ and has significant impact on the the tails of the probability distribution. Thus we focus on the uncorrelated uniform initial condition. In this case
\begin{equation}\label{eq:<e^-pT>=[]^N}
  \overline{ \left\langle e^{-p T}\right\rangle_\mathbf{x} } = 
  \left[
    \int_{-L}^{0} \frac{\dd x}{L} \left\langle e^{-pT}\right\rangle_x
  \right]^N.
\end{equation}
The exact form of $\left\langle e^{-pT}\right\rangle_x$ is given by \eqref{eq:<e^-pT>=int_form} and consists of two terms. Integral of the first term can be computed exactly and its large $L$ behavior reads
\begin{equation}\label{eq:annealed = 1st term}
  \int_{-L}^{0} \frac{\dd x}{L} \erf \left[ \frac{\left|x\right|}{\sqrt{4Dt}}\right]
  \underset{L\to\infty}{\simeq} 1 - \frac{1}{L} \sqrt{ \frac{4 Dt}{\pi} } .
\end{equation}
To determine the behavior of the second term we first integrate with respect to $x$, then take the limit $L\to\infty$ 
\begin{multline}\label{eq:int=1/L int}
  \frac{1}{\pi}
    \int_{-L}^{0} \frac{\dd x}{L} 
    \int_{0}^{1} \dd \xi \frac{ e^{- p t \, \xi} }{\sqrt{\xi(1-\xi)}} 
    \exp \left[ -\frac{x^2}{4Dt}\frac{1}{(1-\xi)} \right]
  \\ 
  =
  \frac{1}{L}
  \sqrt{ \frac{tD}{\pi} }
  \int_{0}^{1}\dd\xi\,
    \frac{e^{- p t \, \xi} }{\sqrt{\xi}}
    \erf\left[ \frac{L}{\sqrt{4Dt(1-\xi)}} \right]
  \\
  \underset{L\to\infty}{\simeq}
    \frac{1}{L}
    \sqrt{ \frac{tD}{\pi} }
    \int_{0}^{1}\dd\xi\,
      \frac{e^{- p t \, \xi} }{\sqrt{\xi}}
  \\=
    \frac{1}{L}
    \sqrt{ \frac{D}{p} }
    \erf\left(\sqrt{pt}\right)
\end{multline} 
Combining \eqref{eq:int=1/L int} with \eqref{eq:annealed = 1st term} we obtain for \eqref{eq:<e^-pT>=[]^N}
\begin{equation}
  \overline{ \left\langle e^{-p T}\right\rangle_\mathbf{x} }
  \underset{L\to\infty}{\simeq}
  \left[
    1 - \frac{1}{L}\sqrt{\frac{4Dt}{\pi}}
    + \frac{1}{L}\sqrt{\frac{D}{p}} \erf \left(\sqrt{p t}\right)
  \right]^{N}
\end{equation}
hence in the thermodynamic limit $N,L\to\infty$ with the fixed ratio $\bar{\rho}=N/L$ we have
\begin{equation}
  \overline{ \left\langle e^{-p T}\right\rangle_\mathbf{x} }
  \underset{L,N\to\infty}{\simeq}
  \exp\left[
    -\bar{\rho} \, \sqrt{4Dt} \; \phi_\text{an}( pt )
  \right]
\end{equation}
where
\begin{equation}\label{eq:phi_an=}
  \phi_\text{an}(q) = \frac{1}{\sqrt{\pi}} - \frac{1}{\sqrt{4q}} \erf\left(\sqrt{q}\right).
\end{equation}
In principle to find the annealed probability distribution $\mathbb{P}_\text{an}[T,t]$ we need to invert double Laplace transform 
\begin{equation}\label{eq:int P_an=DoubleLaplace}
\int_{0^{-}}^{\infty} \dd T\, e^{-pT} \mathbb{P}_\text{an}[T,t] = 
  \exp\left[
    -\bar{\rho} \, \sqrt{4Dt} \; \phi_\text{an}( pt )
  \right].
\end{equation}
However, since  we are interested in the large time behavior, we can instead substitute ansatz \eqref{P_an[T,t]=LD_form} into \eqref{eq:int P_an=DoubleLaplace}. Then after rescaling of the variables we get
\begin{equation}\label{eq:Laplace ansatz substituted}
  \int_{0^-}^{\infty} \dd \tau\, 
    e^{  -\bar{\rho}\sqrt{4Dt}
    \left( \strut q \tau + \Phi_\text{an}(q) \right) } = 
  e^{-\bar{\rho} \, \sqrt{4Dt} \; \phi_\text{an}( q ) },
\end{equation}
where 
\begin{equation}
  q=pt ,
  \qquad
  \tau = \frac{T}{t^{3/2}} \frac{1}{\bar{\rho}\sqrt{4D}}.
\end{equation}
Large time limit corresponds to the saddle point approximation of the integral in \eqref{eq:Laplace ansatz substituted}. By comparing exponents we find that 
\begin{equation}\label{eq:min_tau(qt + Phi_an(t))}
  \min_{\tau}\big( q\tau + \Phi_\text{an}(\tau) \big) = \phi_\text{an}(q).
\end{equation}
Inverting Legendre transform \eqref{eq:min_tau(qt + Phi_an(t))} we obtain
\begin{equation}\label{eq:Phi_an(tau) = max_q}
  \Phi_\text{an}(\tau) = \max_{q}\big( - q\tau + \phi_\text{an}(q) \big) .
\end{equation}
Expression \eqref{eq:Phi_an(tau) = max_q} together with \eqref{eq:phi_an=} provides us with the parametric representation of the annealed large deviation function $\Phi_\text{an}(\tau)$. It is a concave function with the minimum at some point $\tau_\text{typ}$. Due to \eqref{eq:<T>=t^3/2 answer} we indeed expect $\tau_\text{typ} = \frac{1}{3\sqrt{\pi}}$. Furthermore, by analyzing $\phi_\text{an}(q)$ at $q\to0$, $q\to\infty$ and $q\to -\infty$ we can extract asymptotic behavior of $\Phi_\text{an}(\tau)$ for $\tau\to\tau_\text{typ}$, $\tau \to 0$ and $\tau \to \infty$ respectively. 
Below we provide the computation eventually arriving at \eqref{eq:Phi_an(tau)=asympt MainResults}.

\subsubsection{Typical fluctuations}
Typical fluctuations of the occupation time are governed by the behavior of the $\phi_\text{an}(q)$ close to $q=0$. By expanding \eqref{eq:phi_an=} up to the second order in $q$ we find 
\begin{equation}
  \phi_\text{an}(q) \simeq \frac{1}{\sqrt{\pi}} \left( \frac{1}{3} q - \frac{1}{10}q^2 \right),
  \quad q \to 0
\end{equation}
and hence
\begin{equation}
  \Phi_\text{an}(\tau) \simeq \max_q\left( - q \tau 
    + \frac{1}{\sqrt{\pi}} \left( \frac{1}{3} q - \frac{1}{10} q^2 \right) 
  \right).
\end{equation}
This is a quadratic function and we easily find that
\begin{equation}
  \Phi_\text{an}(\tau) \simeq \frac{5}{2} \sqrt{\pi}\left(\tau - \frac{1}{3\sqrt{\pi}}\right)^2, \quad \tau \to \tau_\text{typ} = \frac{1}{3\sqrt{\pi}}
\end{equation}
which implies
\begin{multline}
  \mathbb{P}_\text{an}[T,t] 
    \simeq
  \exp \left[ 
    - \frac{1}{2} 
    \left( \frac{T - T_\text{typ}}{\sigma_\text{an}} \right)^2
  \right], \quad 
  T\to T_\text{typ} \\
\end{multline}
with 
\begin{equation}
  T_\text{typ} = \frac{2}{3} \,\sqrt{\frac{D}{\pi}}  \bar{\rho}\, t^{3/2},
  \qquad
  \sigma_\text{an}^2 = 
  \frac{2}{5}\, \sqrt{\frac{D}{\pi}}\, \bar{\rho}\,t^{5/2}.
\end{equation}
In other words, large deviation function gives us exactly the same results as we have obtained in the previous section by different means (recall \eqref{eq:<T>=t^3/2 answer} and \eqref{eq:Var_an=answer uncorrelated}).

\subsubsection{Atypical fluctuations $T\ll T_\text{typ}$}
Atypically small fluctuations are governed by large $q$ behavior of the $\phi_\text{an}(q)$. Expanding \eqref{eq:phi_an=} in series as $q\to\infty$ we get
\begin{equation}
  \phi_\text{an}(q) \simeq \frac{1}{\sqrt{\pi}} - \frac{1}{\sqrt{4q}},
  \quad q \to \infty.
\end{equation}
Therefore
\begin{equation}
  \Phi_\text{an}(\tau) \simeq \max_{q}\left( -q \tau + \frac{1}{\sqrt{\pi}} - \frac{1}{\sqrt{4q}} \right).
\end{equation}
Maximizing this function we arrive at
\begin{equation}
  \Phi_\text{an}(\tau) \simeq \frac{1}{\sqrt{\pi}} - \frac{3}{2} \left(\frac{\tau}{2}\right)^{1/3} ,
  \quad 
  \tau \to 0.
\end{equation}

\subsubsection{Atypical fluctuations $T\gg T_\text{typ}$ }
To study atypically large fluctuations we need to expand $\phi_\text{an}(q)$ as $q\to-\infty$. The subtlety here lies in the square root, namely we need to perform an analytical continuation of the $\erf(z)$ as
\begin{equation}
   \phi_\text{an}(q) = \frac{1}{\sqrt{\pi}} + 
    \frac{\ii \erf\left( \ii \sqrt{\abs{q}}\right) }{ \sqrt{4\abs{q}}}, \qquad
    q<0,
 \end{equation} 
and using the expansion
\begin{equation}
  \ii \erf( \ii \sqrt{\abs{q}}) \simeq - \frac{e^{\abs{q}}}{\sqrt{\pi \abs{q} }},
  \quad q \to - \infty
\end{equation}
we arrive at
\begin{equation}\label{eq:Phi_an=max t>>t_typ}
  \Phi_\text{an}(\tau) \simeq \max_q\left(  
    \abs{q}\tau + \frac{1}{\sqrt{\pi}} -\frac{1}{\sqrt{\pi}} \frac{e^{\abs{q}}}{2 \abs{q}}
  \right).
\end{equation}
Taking the derivative of the expression in \eqref{eq:Phi_an=max t>>t_typ} we find that its maximum is reached at $q=q^*$ satisfying
\begin{equation}\label{eq:annealed_W-1_eq1}
  \tau - \frac{e^{\abs{q^*}}}{2\sqrt{\pi}\abs{q^*}} = 0
\end{equation}
This is a well-known transcendent equation, and its solution is expressed in terms of the Lambert $W$-function \cite{CGHJK-96}. For $\tau>\frac{e}{2\sqrt{\pi}}$ there are two roots of \eqref{eq:annealed_W-1_eq1}, but since we are looking at $\tau\to\infty$ we should choose the solution corresponding to the larger $\tau$. It is given by the lower branch of the Lambert $W$-function 
\begin{equation}
  q^* = W_{-1}\left( -\frac{1}{2\tau\sqrt{\pi}} \right).
\end{equation}
Utilizing the asymptotic expansion of $W_{-1}(z)$
\begin{equation}\label{eq:Lambert-W -1 expansion}
  W_{-1}(-z) \simeq \log z   - \log \left[ - \log z  \right], \qquad z \to 0
\end{equation} 
we find that in the leading order
\begin{equation}\label{eq:q*=t>>t_typ asymptotic}
  \abs{q^{*}} \simeq \log \left[ 2\tau\sqrt{\pi} \right] + \log \log \left[ 2 \tau\sqrt{\pi} \right], \qquad \tau \to \infty.
\end{equation}
Substituting \eqref{eq:q*=t>>t_typ asymptotic} into \eqref{eq:Phi_an=max t>>t_typ} we get the asymptotic behavior of the large deviation function
\begin{equation}
  \Phi_\text{an}(\tau) \simeq 
  \tau\left( \mu(\tau) \log \left[ 2\tau \sqrt{\pi}\right]  
           - \frac{1}{\mu(\tau)} \right), 
  \quad \tau \to \infty.
\end{equation}
where
\begin{equation}\label{eq:mu(tau)=}
  \mu(\tau) = 1 + \frac{ \log\log \left[ 2\tau \sqrt{\pi} \right]}
                          { \log \left[2\tau\sqrt{\pi}\right] }.
\end{equation}

\subsection{Quenched large deviation function}
Now we proceed to the quenched probability distribution \eqref{eq:P_qu[T,t]=def}. As we have shown in \eqref{eq:P_qu=exp[rho]} its Laplace transform reads 
\begin{equation}\label{eq:P_qu=}
  \int_{0}^{t} \dd T e^{-pT} \mathbb{P}_\text{qu}[T,t] = 
  \exp\left[
    \bar{\rho}
    \int_{0}^{\infty} \dd y
     \log \left\langle e^{-pT}\right\rangle_{y} 
  \right].
\end{equation}
Using the explicit form \eqref{eq:<e^-pT>=int_form} of $\left\langle e^{-pT}\right\rangle_{y}$ we find that the quenched probability distribution is given by 
\begin{equation}
    \int_{0}^{t} \dd T e^{-pT} \mathbb{P}_\text{qu}[T,t] = 
  \exp\left[
    -\bar{\rho}
      \sqrt{4Dt} \; \phi_\text{qu}(pt)
  \right]
\end{equation}
where 
\begin{equation}\label{eq:phi_qu=}
  \phi_\text{qu}(q) = 
    -\int_{0}^{\infty} \dd y
     \log \left[
      \erf(y) 
      + 
      \frac{1}{\pi} \int_{0}^{1} \dd \xi \frac{ e^{- q \xi - \frac{y^2}{1-\xi}} }{\sqrt{\xi(1-\xi)}} 
     \right] . 
\end{equation}
Analogously to the annealed case we find that in the saddle-point approximation 
\begin{equation}
  \mathbb{P}_\text{qu}[T,t] \simeq \exp \left[  - \bar{\rho} \sqrt{4Dt}\;\Phi_\text{qu}(\tau) \right], \quad
  \tau = \frac{T}{t^{3/2}} \frac{1}{\bar{\rho}\sqrt{4D}},
\end{equation}
where the large deviation function $\Phi_\text{qu}(\tau)$ is given by an inverse Legendre transform
\begin{equation}\label{eq:PHI_qu=max_q}
  \Phi_\text{qu}(\tau) = \max_q \left( -q\tau  + \phi_\text{qu}(q)\right).
\end{equation}
Equations \eqref{eq:PHI_qu=max_q} and \eqref{eq:phi_qu=} give the parametric representation of $\Phi_\text{qu}(\tau)$. Let us now study its the asymptotic behavior and derive \eqref{eq:Phi_qu(tau)=asympt MainResults}.

\subsubsection{Typical fluctuations}
In order to obtain fluctuations of the occupation time close to the typical value we need to study the behavior of $\phi_\text{qu}(q)$ as $q\to0$. By first expanding the integrals in \eqref{eq:phi_qu=} in series with respect to $q$ and then  using the representation of the moments in terms of confluent hypergeometric function \eqref{eq:con_hg_function=def} as in \eqref{eq:<T>=Psi(..,..,..)}, after technical yet straightforward computation we arrive at
\begin{equation}
  \phi_\text{qu}(q) \simeq \frac{1}{3\sqrt{\pi}} q - \frac{4}{15} \frac{\sqrt{2}-1}{\sqrt{\pi}} q^2, \quad q \to 0.
\end{equation}
Therefore in accordance with \eqref{eq:PHI_qu=max_q} we have
\begin{equation}
  \Phi_\text{qu}(\tau) \simeq \max_{q}\left( -q\tau + \frac{1}{3\sqrt{\pi}}q - \frac{4}{15}\frac{\sqrt{2}-1}{\sqrt{\pi}}q^2 \right).
\end{equation}
Maximizing this quadratic function yields
\begin{equation}\label{eq:Phi_qu(tau)= tau->tau_typ}
  \Phi_\text{qu}(\tau) \simeq 
  \frac{1}{2}\frac{15\sqrt{\pi}}{4(\sqrt{2}-1)} \left(\tau - \frac{1}{3\sqrt{\pi}}\right)^{2}, \quad \tau \to \tau_\text{typ} = \frac{1}{3\sqrt{\pi}}.
\end{equation}
Expression \eqref{eq:Phi_qu(tau)= tau->tau_typ} implies that close to the typical value, the quenched probability distribution behaves as
\begin{equation}
  \mathbb{P}_\text{qu}[T,t] \simeq \exp\left[ - \frac{1}{2} \left( \frac{T - T_\text{typ}}{\sigma_\text{qu}}\right)^2 \right], \quad 
  T \to T_\text{typ}
\end{equation}
where
\begin{equation}
  T_\text{typ} = \frac{2}{3} \,\sqrt{\frac{D}{\pi}}  \bar{\rho}\, t^{3/2},
  \quad
  \sigma_\text{qu}^2 = \frac{8 \left(\sqrt{2}-1\right) }{15} \sqrt{\frac{D}{\pi}}\, \bar{\rho} \, t^{5/2}. 
\end{equation}
This is exactly the behavior anticipated in \eqref{eq:<T>=t^3/2 answer} and \eqref{eq:Var_qu=ans t^5/2}.

\subsubsection{Atypical fluctuations $T\ll T_\text{typ}$}
Atypically small fluctuations can be extracted from the $q\to\infty$ expansion of $\phi_\text{qu}(q)$. First we note, that in \eqref{eq:phi_qu=} the main contribution to the integral over $\xi$ comes from values of $\xi$ close to zero, therefore
\begin{equation}\label{eq:large q int dxi 1}
  \int_{0}^{1} \dd \xi\, \frac{ e^{- q \xi - \frac{y^2}{1-\xi}} }{\sqrt{\xi(1-\xi)}} 
  \simeq
  e^{-y^2}\int_{0}^{1}\dd\xi\, \frac{ e^{-q\xi} }{\sqrt{\xi}},
  \quad 
  q\to\infty.
\end{equation}
Computing this integral yields
\begin{equation}\label{eq:large q int dxi 2}
  \int_{0}^{1}\dd\xi\, \frac{ e^{-q\xi } }{\sqrt{\xi}}  
  =
  \frac{\sqrt{\pi}\erf \left( \sqrt{q} \right)}{\sqrt{q}}
  \simeq \sqrt{\frac{\pi}{q}},
  \quad q\to\infty.
\end{equation}
Substituting \eqref{eq:large q int dxi 2} and \eqref{eq:large q int dxi 1} into \eqref{eq:phi_qu=} we arrive at
\begin{equation}\label{eq:phi_qu=large q asympt1}
  \phi_\text{qu}(q) \simeq 
    - \int_{0}^{\infty} \dd y\, 
      \log\left[ \erf (y) + e^{-y^2} \frac{1}{\sqrt{\pi q}} \right]
  , \quad q\to\infty.
\end{equation}
To proceed further we use a trick. Namely, we take a derivative of $\phi_\text{qu}(q)$
\begin{equation}\label{eq:d/dq phi_q}
  \pdv{}{q}\phi_\text{qu}(q) = \frac{1}{2q^{3/2}} \int_{0}^{\infty} \dd y\, 
  \frac{1}{\frac{1}{\sqrt{q}} + e^{y^2}\sqrt{\pi} \erf (y)}.
\end{equation}
and note, that the integral \eqref{eq:d/dq phi_q} diverges as $q\to\infty$. To determine the large $q$ behavior of $\phi_\text{qu}(q)$, it is sufficient to find the divergent part of \eqref{eq:d/dq phi_q} and integrate it back. This can be done by rewriting the integral in \eqref{eq:d/dq phi_q} as 
\begin{multline}\label{eq:d/dq phi_q = int + int + int + int}
  \int_{0}^{\infty} \dd y\, 
  \frac{1}{\frac{1}{\sqrt{q}} + e^{y^2}\sqrt{\pi} \erf y} 
    \\
  = 
  \int_{0}^{1} \dd y
  \left(
    \frac{1}{\frac{1}{\sqrt{q}} + e^{y^2}\sqrt{\pi} \erf (y)}
    -
    \frac{1}{\frac{1}{\sqrt{q}} + 2 y}
  \right) 
  \\
  +
  \int_{1}^{\infty} \dd y\, 
  \frac{1}{\frac{1}{\sqrt{q}} + e^{y^2}\sqrt{\pi} \erf (y)} 
  +
  \int_{0}^{1}\frac{1}{\frac{1}{\sqrt{q}} + 2 y} \dd y.
\end{multline}
In \eqref{eq:d/dq phi_q = int + int + int + int} first two integrals converge, computing them we get
\begin{multline}
    \int_{0}^{\infty} \dd y\, 
  \frac{1}{\frac{1}{\sqrt{q}} + e^{y^2}\sqrt{\pi} \erf (y)} 
  \\
  \simeq
  \frac{1}{4}\log \frac{\pi}{4}
  +
  \int_{0}^{1}\frac{1}{\frac{1}{\sqrt{q}} + 2 y} \dd y
\end{multline}
the remaining integral diverges logarithmically, hence 
\begin{equation}\label{eq:pdv phi_qu ~ log}
  \pdv{}{q}\phi_\text{qu}(q) \simeq \frac{1}{8 q^{3/2}} \log \pi q,
  \quad q\to\infty .
\end{equation}
Integrating \eqref{eq:pdv phi_qu ~ log} back gives us
\begin{equation}
  \phi_\text{qu}(q) = \phi_\infty - \frac{2 + \log \pi q}{4\sqrt{q}},
\end{equation}
where the constant of integration is restored by taking the limit $q\to\infty$ in \eqref{eq:phi_qu=large q asympt1}
\begin{equation}
  \phi_\infty = - \int_{0}^{\infty} \log\left[ \erf(y) \right] \dd y \approx 1{.}03.
\end{equation}
The only thing left is to invert the Legendre transform 
\begin{equation}
  \Phi_\text{qu}(\tau) \simeq \max_q\left( -q\tau + \phi_\infty - \frac{2 + \log \pi q}{4\sqrt{q}} \right)
\end{equation}
Maximum of this expression is reached at $q^{*}$, a root of
\begin{equation}\label{eq:d/dq phi_qu = 0 large q}
  -\tau + \frac{1}{8q^{3/2}} \log \pi q^* = 0.
\end{equation}
After a change of variables $q = \frac{1}{\pi}\exp\left[ -\frac{2}{3}u\right]$ equation \eqref{eq:d/dq phi_qu = 0 large q} transforms into
\begin{equation}
  e^u + \frac{12 \tau  }{ u\, \pi^{3/2}} = 0.
\end{equation}
We have already solved this equation in \eqref{eq:annealed_W-1_eq1} and its solution is given in terms of the lower branch of the Lambert $W$-function as
\begin{equation}
  q^* = \frac{1}{\pi} \exp\left[ - \frac{2}{3} \, W_{-1}\left( -\frac{12\tau}{\pi^{3/2}} \right) \right], \qquad  \tau \to 0.
\end{equation}
Using the asymptotic expansion for $W_{-1}(z)$ as in \eqref{eq:Lambert-W -1 expansion} we find that in the leading order
\begin{equation}
  q^{*} \simeq \left(\frac{1}{12\tau}\log \frac{\pi^{3/2}}{12\tau} \right)^{2/3}, \qquad \tau \to 0
\end{equation}
and hence 
\begin{equation}
  \Phi_\text{qu}(\tau) \simeq
  \phi_{\infty} - 
  \frac{ 1 + \frac{1}{2} \log \nu  + \frac{1}{3}\log\log \nu}
       { \frac{2}{\sqrt{\pi}}\left(\nu \log \nu\right)^{1/3} },
  \quad \tau \to 0
\end{equation}
where
\begin{equation}
  \nu = \frac{\pi^{3/2}}{12\tau}.
\end{equation}

\subsubsection{Atypical fluctuations $T\gg T_\text{typ}$}
Atypically large fluctuations correspond to $q\to-\infty$ expansion of $\phi_\text{qu}(q)$. To obtain this expansion we perform the change of variables $y \mapsto y\sqrt{\abs{q}}$ and $\xi\mapsto1-\xi$ in \eqref{eq:phi_qu=} arriving at
\begin{multline}\label{eq:phi_qu = q->-infty}
  \phi_\text{qu}(q) = 
  -\sqrt{\abs{q}}
  \int_{0}^{\infty} \dd y\, 
  \\
  \log\left[
      \erf\left( y \sqrt{\abs{q}} \right)
      + 
      \frac{1}{\pi}
      \int_{0}^1 \dd \xi\;
      \frac{e^{ -\abs{q}\left( \frac{y^2}{\xi} + \xi -1 \right) } }
           {\sqrt{\xi(1-\xi)}}
  \right].
\end{multline}
The integral over $\xi$ can be evaluated in the saddle-point approximation yielding
\begin{equation}
  \frac{1}{\pi}
      \int_{0}^1 \dd \xi\;
      \frac{e^{ -\abs{q}\left( \frac{y^2}{\xi} + \xi -1 \right) } }
           {\sqrt{\xi(1-\xi)}}
  \simeq \frac{ e^{ -\abs{q}(2y-1) } }{\sqrt{\pi\abs{q}(1-y)}},
  \quad q \to -\infty,
\end{equation}
hence we have
\begin{equation}
  \phi_\text{qu}(q) 
    \simeq
  -\sqrt{\abs{q}}
  \int_{0}^{\infty} \dd y\, 
  \log\left[
      1 + \frac{ e^{ -\abs{q}(2y-1) } }{\sqrt{\pi\abs{q}(1-y)}}
  \right].
\end{equation}
Integrating by parts we get
\begin{equation}
  \phi_\text{qu}(q) \simeq
  -\sqrt{\abs{q}} \int_{0}^{1} \dd y\; 
    \frac{2y\abs{q}}{ 1 + \sqrt{\pi\abs{q}} \, e^{-\abs{q}(1 - 2y)} }.
\end{equation}
This is a complete Fermi-Dirac integral which can be evaluated in terms of the polylogarithm function 
\begin{equation}
  \phi_\text{qu}(q) \simeq \frac{1}{2\sqrt{\abs{q}}} \mathrm{Li}_2\left( -\frac{e^{\abs{q}}}{\sqrt{\pi\abs{q}}} \right)
  ,\quad q \to -\infty
\end{equation}
and using the expansion 
\begin{equation}
  \mathrm{Li}_2(-z)\simeq - \frac{1}{2} \log^2(z),
  \quad z \to \infty
\end{equation}
we finally find that
\begin{equation}
  \phi_\text{qu}(q)\simeq - \frac{1}{4}\abs{q}^{3/2},\quad 
q\to-\infty.
\end{equation}
Thus the large deviation function is given by
\begin{equation}
  \Phi_\text{qu}(\tau) \simeq \max_q\left( -q \tau - \frac{1}{4}(-q)^{3/2} \right)
\end{equation}
and hence
\begin{equation}
  \Phi_\text{qu}(\tau) \simeq \left( \frac{4}{3}\,\tau \right)^{3}, \quad 
  \tau\to \infty.
\end{equation}

\section{Numerical simulations}\label{sec:numerics}
\par To support our analytical calculations we perform numerical simulations. 
Since there is no interaction, the total occupation time is nothing but a sum of occupation times for individual particles. Therefore, in order to explore probability distributions \eqref{eq:P_an[T,t]=def} and \eqref{eq:P_qu[T,t]=def} we can sample $N$ single particle occupation times from \eqref{eq:P[T,t|x]=1p} while choosing appropriate initial conditions. However in this process we encounter two problems. 

\par The first problem involves effectively sampling from \eqref{eq:P[T,t|x]=1p}.
The second problem is more intricate, as it entails capturing samples with atypical values of the occupation time. Since the probability of getting such values is very small, we cannot reach the tails of the distribution when sampling from \eqref{eq:P[T,t|x]=1p}.
Fortunately, both problems can be resolved. To address the first problem, we introduce an additional variable corresponding to the first hitting time and sample occupation time in two steps. The second problem, although more subtle, is well known and in many situations can be resolved by Importance Sampling Monte-Carlo \cite{H-02,H-11,NMV-11,HDMRS-18,BMRS-19,BMR-23}. Here we shall utilize this approach as well.

\subsection{Sampling strategy}
\par First we address the problem of sampling from the probability distribution of the single particle occupation time $\mathbb{P}[T,t,\vert\,x]$. To do this we rewrite \eqref{eq:P[T,t|x]=1p} as 
\begin{multline}\label{eq:P[T,t|x] = int F[w]}
  \mathbb{P}[T,t\,\vert\, x] = 
  \int_{t}^{\infty} \dd \omega \,
    \delta(T) \; \mathbb{F}[\omega\,\vert\, x] 
  \\ +
  \int_{0}^{t} \dd \omega \,
    \mathbb{P}[T, t-\omega\, \vert\, x = 0] \;
    \mathbb{F}[\omega \, \vert\, x]
\end{multline}
where 
\begin{equation}\label{eq:F[w|x]=def}
  \mathbb{F}[\omega \, \vert\, x] = 
    \frac{\abs{x}}{\sqrt{4\pi D \omega^{3}}}
    \exp\left[ - \frac{x^2}{4D \omega} \right].
\end{equation}
Function $\mathbb{F}[\omega \, \vert\, x]$ in \eqref{eq:F[w|x]=def} is nothing but a probability distribution of the first passage time \cite{Redner-first-passage}, i.e. $\omega$ is the time at which Brownian particle reaches the origin for the first time given the initial position $x$. 
The factor $\mathbb{P}[T,t-\omega\,\vert\,x=0]$ according to \eqref{eq:P[T,t|x]=1p} is 
\begin{equation}\label{eq:P[T,t-w,0]=}
  \mathbb{P}[T,t-\omega \, \vert\, x = 0] = 
  \frac{1}{\pi} \frac{1}{\sqrt{T(t-\omega - T)}} \,
  \mathds{1}_{[0,t-\omega]}(T).
\end{equation}
In \eqref{eq:P[T,t-w,0]=} we added an indicator function, to emphasize, that the distribution is supported on the segment $[0,t-\omega]$.

\par The sampling procedure for the occupation time of an individual particle is now as follows: first we sample $\omega$  from $\mathbb{F}[\omega\,\vert\, x]$ given by \eqref{eq:F[w|x]=def}; if $\omega\ge t$, i.e. the particle does not reach the origin up to the observation time, then occupation time is zero (it corresponds to the first term in \eqref{eq:P[T,t|x] = int F[w]}), otherwise we sample occupation time $T$ from $\mathbb{P}[T,t-\omega \, \vert\, x = 0] $ in \eqref{eq:P[T,t-w,0]=}. Since both \eqref{eq:F[w|x]=def} and \eqref{eq:P[T,t-w,0]=} are relatively simple, we sample from them directly. This way, we can effectively draw occupation time for a single particle with  fixed initial position.

\par Now we proceed to the problem of exploring the tails of the probability distribution. To address it, we utilize Importance Sampling Monte-Carlo technique. Let us briefly recall its basics.

\par Consider a random variable $z$ with probability distribution $\mathbb{P}[z]$ and suppose that our goal is to compute an average value of some observable $O(z)$
\begin{equation}\label{eq:<O>=def}
  \left\langle O(z)\right\rangle = 
  \int\dd z\, O(z) \mathbb{P}[z]
\end{equation}
The usual Monte-Carlo strategy is to get $n$ samples $z_i$ from $\mathbb{P}[z]$ and estimate an average by
\begin{equation}\label{eq:<O>=MonteCarlo}
  \left\langle O(z) \right\rangle \approx \frac{1}{n} \sum_{i} O(z_i),
  \qquad z_i \leftarrow  \mathbb{P}[z].
\end{equation}
To study the probability distribution itself we can choose the observable to be an indicator function
\begin{equation}
  O(z) = \mathds{1}_{[z_1; z_2]}(z) \equiv 
  \begin{cases}
    1, & z \in [z_1,z_2],\\
    0, & z \notin [z_1, z_2].
  \end{cases}
\end{equation}
The average value of the indicator function is nothing but the probability that $z$ falls into the interval $[z_1,z_2]$
\begin{equation}
  \left\langle \mathds{1}_{[z_1; z_2]}(z) \right\rangle 
    = \mathbb{P}[z_1 \le z \le z_2] .
\end{equation}
By choosing sufficiently small interval $[z_0, z_0 + \dd z]$ we get the probability distribution itself
\begin{equation}
  \left\langle \mathds{1}_{[z_0,z_0+\dd z]}(z) \right\rangle = \mathbb{P}[z_0]\, \dd z. 
\end{equation}
If we are interested in the tails of the probability distribution where $\mathbb{P}[z]$ is usually very small, then the above procedure is ill-suited. Indeed, the smaller $\mathbb{P}[z_0]$ the rarer we get values laying in $[z_0, z_0 + \dd z]$. Of course, in the limit $n\to\infty$ approximation  \eqref{eq:<O>=MonteCarlo} becomes exact, however in practice it may require unreasonably many samples.

\par To overcome this problem we rewrite \eqref{eq:<O>=def} as 
\begin{equation}\label{eq:<O> = int O P/Q}
  \left\langle O(z) \right\rangle = 
  \int\dd z\, \left( O(z) \frac{\mathbb{P}[z]}{\mathbb{Q}[z]} \right) \mathbb{Q}[z]
\end{equation}  
where $\mathbb{Q}[z]$ is some, for the moment arbitrary, probability distribution. Comparing \eqref{eq:<O> = int O P/Q} with \eqref{eq:<O>=def} we see that instead
of \eqref{eq:<O>=MonteCarlo} we can use 
\begin{equation}\label{eq:<O>=MonteCarlo_IS}
  \left\langle O(z) \right\rangle \approx 
  \frac{1}{n} \sum_{i} O(z_i) \frac{\mathbb{P}[z_i]}{\mathbb{Q}[z_i]},
  \qquad z_i \leftarrow  \mathbb{Q}[z].
\end{equation}
where $z_i$ are drawn from $\mathbb{Q}[z]$ and not from $\mathbb{P}[z]$ as in \eqref{eq:<O>=MonteCarlo}. 

\par 
Formally \eqref{eq:<O>=MonteCarlo} and \eqref{eq:<O>=MonteCarlo_IS} are indeed equivalent. But by an appropriate choice of $\mathbb{Q}[z]$ we can drastically decrease the number of samples required for an approximation to be accurate enough.

\par Now we return to the problem of a Brownian motion. To implement Importance Sampling strategy, we introduce an exponential tilt in \eqref{eq:P[T,t|x]=1p}, and sample occupation time from
\begin{equation}\label{eq:Q[T,t|x]}
  \mathbb{Q}[T,t\,\vert\,x] = \frac{1}{Z(\beta, t\,\vert\,x)}e^{ -\beta \frac{T}{t} } \mathbb{P}[T,t\,\vert\,x]
\end{equation}
where the normalization $Z(\beta,t\,\vert\,x)$ is
\begin{equation}\label{eq:Z(beta,t|x)=}
  Z(\beta,t\,\vert\,x) = \int_{0}^{t} \dd T\,  e^{ - \beta \frac{T}{t} } \mathbb{P}[T,t\,\vert\,x].
\end{equation}
By varying the parameter $\beta$ in  \eqref{eq:Q[T,t|x]} we can bias the trajectories. Namely, positive (negative) values of $\beta$ bias the trajectories with small (large) occupation time. To get the sample of the occupation time from the tilted distribution \eqref{eq:Q[T,t|x]} we use Metropolis algorithm \cite{Krauth-06}, i.e. sample new value of occupation time $\tilde{T}$ from $\mathbb{P}[T,t\,\vert\,x]$ and accept the ``move'' with probability $\min\left(1, e^{ -\beta \frac{\tilde{T}-T}{t} }\right)$.

\subsection{Probability distributions}
\begin{figure*}
\includegraphics[width = \linewidth]{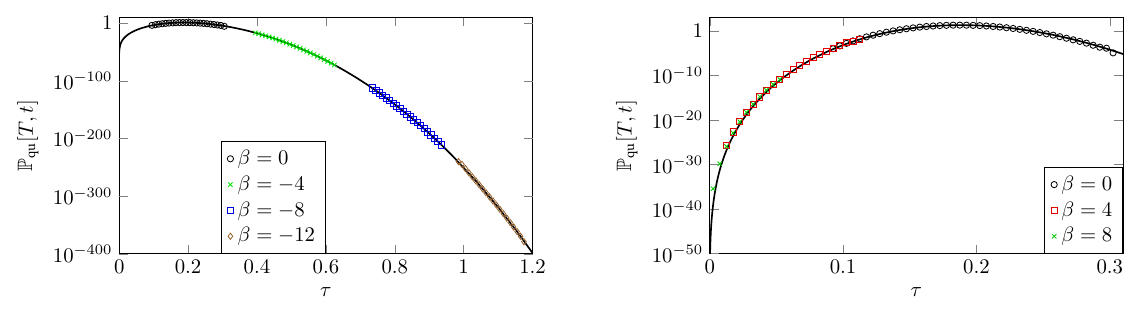}
\caption{The quenched probability distribution obtained from \eqref{eq:P_qu[T,t]=LD form MainResults}, \eqref{eq:Phi_qu(tau)=max() MainResults} and \eqref{eq:phi_qu(tau)=MainResults} (solid lines). Results of the simulations for the negative (positive) values of tilt $\beta$ in \eqref{eq:Q[T,t|x]} are shown in the left (right) plane.  Different shapes correspond to the different values of the tilt $\beta$.}
\label{fig:numerics_quenched}
\end{figure*}
\begin{figure*}
\includegraphics[width = \linewidth]{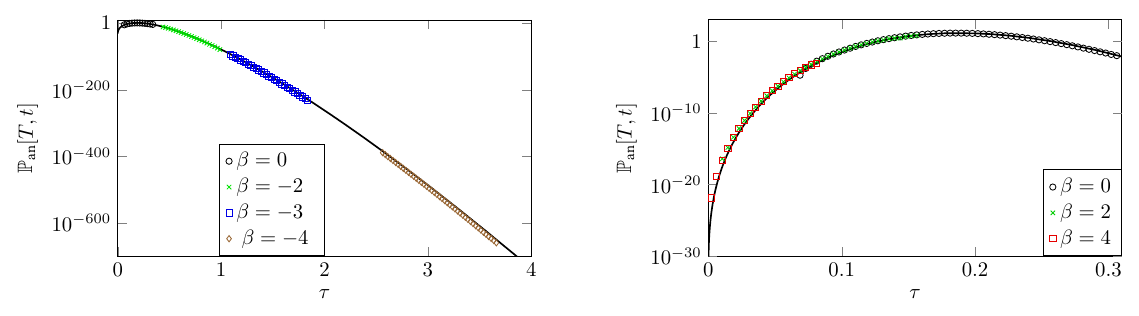}
\caption{The annealed probability distribution obtained from \eqref{eq:P_an[T,t]=LD form MainResults}, \eqref{eq:Phi_an(tau)=max() MainResults} and \eqref{eq:phi_an(tau)= MainResults} (solid lines). Results of the simulations for the negative (positive) values of tilt $\beta$ in \eqref{Q[T,t;x]=} are shown in the left (right) plane.  Different shapes correspond to the different values of the tilt $\beta$.}
\label{fig:numerics_annealed}
\end{figure*}

\subsubsection{Quenched distribution}
The quenched probability distribution corresponds to the typical configuration of particles. 
One of the ways to realize a typical configuration is to take equidistantly distributed particles. 
Therefore to mimic the quenched probability distribution we consider frozen initial condition (example (a) in Fig.~\ref{fig:initial_conf})
\begin{equation}\label{eq:x_i=quenched}
  x_i \equiv x_{i}(0) = - \frac{i - \frac{1}{2}}{\bar{\rho}},
  \qquad i = 1,\ldots,N.
\end{equation}

\par The quenched probability distribution can then be approximated as
\begin{equation}
  \mathbb{P}_\text{qu}[T,t] \dd T \approx 
  \frac{1}{n}
  \sum_{i} \mathds{1}_{[T,T+\dd T]}\left(\textstyle\sum_{j=1}^{N} T_j  \right)
  \prod_{j=1}^{N}
  \frac{\mathbb{P}[T,t \,\vert\, x_j]}
       {\mathbb{Q}[T,t \,\vert\, x_j]}
\end{equation}
with $x_j$ given by \eqref{eq:x_i=quenched} and $T_j$ sampled from the tilted distribution $\mathbb{Q}[T,t\,\vert\,x_i]$ as in \eqref{eq:Q[T,t|x]}
\begin{equation}
  T_j \leftarrow \mathbb{Q}[T,t\,\vert\,x_j].
\end{equation}
To get the sample of $N$ single particle occupation times, we use Metropolis algorithm. At each step we pick some fraction $r < N$ of single particle occupation times $T_j$ and resample them. The value of $r$ is chosen in such a way that approximately half of proposed Metropolis steps are accepted.  

\par In numerical simulation we use $D=0{.}5$, $N=10^{4}$, $L=10^{4}$ ($\bar{\rho}=1$) and $t=1000$. For each value of $\beta$ we produce $10^{6}$ configurations (that is $10^{10}$ samples of a single particle occupation times). To compare simulations with the analytical results we compute $Z(\beta,t\,\vert\,x_j)$ in \eqref{eq:Z(beta,t|x)=} and the normalization constant in \eqref{P_qu[T,t]=LD_form} numerically. 
Resulting plots are given in Fig~\ref{fig:numerics_quenched}.

\subsubsection{Annealed distribution}

\par The only difference with respect to the quenched case is that now initial coordinates are not fixed but drawn from the uniform distribution on the segment $[-L;0]$. Hence we approximate the probability distribution as   
\begin{equation}\label{eq:annealed IS monte-carlo}
  \mathbb{P}_\text{an}[T,t] \dd T \approx 
  \frac{1}{n}
  \sum_{i} \mathds{1}_{[T,T+\dd T]}\left(\textstyle\sum_{j=1}^{N} T_j  \right)
  \prod_{j=1}^{N}
  \frac{\mathbb{P}[T_j,t ; x_j]}
       {\mathbb{Q}[T_j,t ; x_j]}.
\end{equation}
where we sample $(T_j,x_j)$ from
\begin{equation}
  (x_j, T_j) \leftarrow \mathbb{Q}[T,t; x_j].
\end{equation}
and $\mathbb{P}[T,t;x]$ and $\mathbb{Q}[T,t;x]$ are joint distributions of the occupation time and initial coordinate, given by
\begin{align}
  \label{P[T,t;x]=}
  & \mathbb{P}[T,t;x] = \frac{1}{L}\, \mathbb{P}[T,t\,\vert\,x],
  \\ 
  \label{Q[T,t;x]=}
  & \mathbb{Q}[T,t;x] = \frac{1}{Z(\beta,t)}e^{-\beta \frac{T}{t}} \mathbb{P}[T,t;x].
\end{align}
The normalization in \eqref{Q[T,t;x]=} is
\begin{equation}\label{eq:Z(beta,t)=}
  Z(\beta,t) = \int_{-L}^{0} \dd x \int_{0}^{t} \dd T\,
    e^{-\beta \frac{T}{t} }\mathbb{P}[T,t;x].
\end{equation}
Note that constants in \eqref{Q[T,t;x]=} and \eqref{eq:Q[T,t|x]} are related via
\begin{equation}
  Z(\beta,t) = \int_{-L}^{0} \frac{\dd x}{L} Z(\beta,t\,\vert\,x).
\end{equation}
The sampling strategy is the same as in the quenched case, and the only difference is that we instead of using \eqref{eq:x_i=quenched} we sample $x$ from the uniform distribution.

\begin{figure*}
\includegraphics[width = \linewidth]{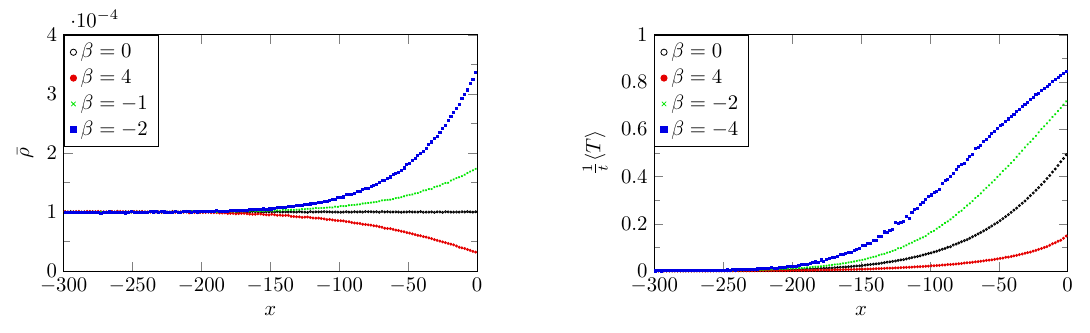}
\caption{The annealed case: Average initial density profile (left) and average value of the single particle occupation time as a function of its initial position (right). Different shapes correspond to the different values of the tilt $\beta$. From the left plane it is clear that atypically large values of the occupation time correspond to the initial conditions with particles localized close to the origin whereas atypically small values of the occupation time correspond to the configuration with the vicinity of the origin being depopulated. Right plane shows that, as in the quenched case, atypical values of the occupation time are due the particles initially located close to the origin.}
\label{fig:numerics_an_average_density}
\end{figure*}

\par In numerical simulation we use $D=0{.}5$, $N=10^{4}$, $L=10^{4}$ ($\bar{\rho}=1$) and $t=1000$. For each value of $\beta$ we produce $10^{6}$ configurations (that is $10^{10}$ samples of a single particle occupation times). To compare numerical results with the analytical we compute $Z(\beta,t)$ in \eqref{eq:Z(beta,t)=} and the normalization constant in \eqref{P_an[T,t]=LD_form} numerically. Resulting plots are given in Fig~\ref{fig:numerics_annealed}.

\par As a side comment, let us mention, that on the contrary to \cite{BMR-23} we do not directly bias the distribution of the initial coordinates of particles.

\begin{figure}
\includegraphics[width = \linewidth]{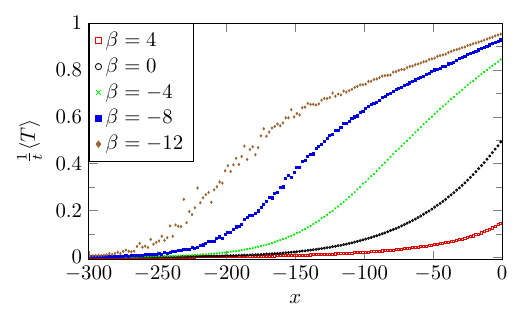}
\caption{The quenched case: Average value of the single particle occupation time as a function of its initial position. Different shapes correspond to the different values of the tilt $\beta$. Atypically small and atypically large values of the total occupation time originate from a bunch of particles close to the origin.}
\label{fig:numerics_qu_dominating_configurations}
\end{figure}

\subsection{Impact of the initial conditions}

\par One of the advantages of the Metropolis algorithm is that it tells us which trajectories contribute to the atypical values of the occupation time.

\par In the quenched case, the initial coordinates of particles are fixed, and atypically large values in principle can be reached via two mechanisms. First option is that a lot of particles contribute slightly more than they typically would, but due to their numbers, total occupation time is atypically large.  
The second option is that the main contribution comes from a small number of particles which have extremely large values of the occupation time. 
By plotting average values of the single particle occupation times $T_j$ (Fig~\ref{fig:numerics_qu_dominating_configurations}) we see that the latter mechanism is in work here. For the atypically small values of the occupation time we see a similar picture. In other words both tails of the quenched probability distributions are governed by just a handful of particles. 

\par As a consistency check, recall that the mean square displacement of a Brownian particle is $\sqrt{\left\langle \Delta x^2\right\rangle} = \sqrt{2Dt}$ which in our simulation is $\sqrt{\left\langle \Delta x^2 \right\rangle} = 100$. Therefore we expect approximately $100$ out of $10000$ particles to contribute to the typical values of the occupation time. From Fig.~\ref{fig:numerics_qu_dominating_configurations} we see that this is exactly the case.

\par On the contrary to the quenched distribution, the annealed one takes into account atypical initial conditions. As we argued in Sec~\ref{seq:results}, this leads to the higher probabilities of both atypically large and atypically small values of the occupation times. Indeed, configurations with particles initialized close to (far from) the origin should lead to large (small) values of the occupation time. This can be observed by plotting average initial density of particles for different values of the tilt $\beta$ (see Fig.~\ref{fig:numerics_an_average_density}, left). In addition, by plotting average value of the occupation time as a function of the initial position (see Fig.~\ref{fig:numerics_an_average_density}, right), we see that, as in the quenched case, atypical values of the occupation time are governed by a small faction of particles located close to the origin. 
This clearly shows that both randomness in the initial configuration and stochasticity of trajectories contribute to the atypical values of the occupation time in the annealed case.

\section{Conclusion}\label{sec:conclusion}

\par We have considered a system of noninteracting Brownian particles on the line with steplike initial conditions focusing on a particular observable, occupation time on the half line. In this paper we have derived several exact asymptotic results that are summarized below.

\par 
We have computed the mean and the variance of the occupation time on the half-line for the general steplike initial condition. 
We have shown that the dependence of the variance on the initial condition is encoded in the single static quantity known as generalized compressibility (Fano factor). 
For the uncorrelated uniform initial condition we have found the large deviation forms of the quenched and the annealed probability distributions.   
By computing the corresponding large deviation functions we have described the tails of these probability distributions. 
Two resulting distributions are very different. 
As a verification of the theoretical results we performed numerical simulations with Importance Sampling Monte-Carlo. 

\par 
We emphasize that, despite the similarity in the methods, the results for the occupation time presented in this paper could have not been predicted based on those  obtained by the authors in \cite{BMR-23} for the local time. 
However, given the similarity in the analysis, it is natural to inquire whether the same methods can be extended further to study other functionals, such as the area under a Brownian excursion, for example.

\par  
There are several natural directions in which it would be interesting to generalize our results.  
First and foremost we can consider more general, though still noninteracting systems. This includes introducing external potential or stochastic resetting, or replacing the Brownian motion with general Stationary process \cite{NT-18-PRL,B-19},  L\'evy flights, fractional Brownian motion, continuous random walks \cite{WBB-20,BB-20,SWSWKCM-21} or active run-and-tumble particles. 
Another direction is adding an interaction into the system, and consider, for example noncrossing particles \cite{MS-23}, ranked diffusion \cite{D-21,FDMS-23} or Brownian particles with annihilation or coalescence \cite{KB-15}. 
Finally it would be interesting to step out of one dimension and ask all the questions mentioned above in two or three dimensions.

\begin{appendix}

\section{Comparison with the Local Time Density}\label{sec:app-OTvsLTD}

\par As mentioned previously, the analysis we conducted in this paper to study the occupation time on the positive half-line is, in fact, very similar to the one performed in \cite{BMR-23} for another Brownian functional: the local time density at the origin. Therefore, it is natural to present the results side by side and compare them, highlighting the similarities and differences between them. Below, we provide such a discussion.

\par Since we have used the notation $T$ to denote both the occupation time on the positive half-line (in this paper) and the local time density at the origin (in \cite{BMR-23}), to prevent confusion while comparing the results, we shall denote them by $T^\text{occ}$ and $T^\text{loc}$ respectively.

\par Both of the aforementioned observables belong to the class of additive Brownian functionals, and have the form  
\begin{equation} 
  O = \sum_{i=1}^{N}\int_{0}^{t} V\left[ x_i(t')\right] \dd t'
\end{equation}
The difference between $T^\text{occ}$ and $T^\text{loc}$ lies in the choice of the functional $V[x(t)]$. Specifically, the occupation time on the positive half-line corresponds to $V[x(t)] = \theta[x(t)]$ whereas for the  local time density $V[x(t)] = \delta[x(t)]$ 
\begin{align}
  \label{eq:app occ time = def}
  & T^\text{occ} \equiv \sum_{i} \int_{0}^{t} \theta[x_i(t')] \, \dd t' ,
  \\
  \label{eq:app loc time = def}
  & T^\text{loc} \equiv \sum_{i} \int_{0}^{t} \delta[x_i(t')] \, \dd t' .
\end{align} 
\par The occupation time quantifies how long particles reside in the positive half-line an it is exactly the \emph{time} that the particles spent to the right of the origin.  Since $\theta[x]$ is either $0$ or $1$, from \eqref{eq:app occ time = def} it is evident, that if there are $N$ particles in the system, then $T^\text{occ}$ cannot be greater than $Nt$. In other words, its probability distribution is supported on the segment $[0,Nt]$. 
\begin{equation}
  \dim \left[ T^{\text{occ}} \right] =  \text{time},
  \qquad
  T^{\text{occ}} \in [0,Nt].
\end{equation}

\par The quantity $T^{\text{loc}}$ in \eqref{eq:app loc time = def}, on the other hand, characterizes the amount of time particles have spent in the vicinity of the origin, but in a slightly more subtle way, as it is \emph{time density} rather than time itself. Consequently, and this can be easily seen from \eqref{eq:app loc time = def},  the local time density is not bounded and can take arbitrarily large values. 
\begin{equation}
  \dim \left[ T^{\text{loc}} \right] =  \frac{\text{time}}{\text{length}},
  \qquad
  T^{\text{loc}} \in [0,\infty).
\end{equation}
\par At first glance, these two observables appear quite similar. Indeed, since $\pd_x\theta(x) = \delta(x)$, they are related on the functional level. However, the reader should not let this similarity mislead them, since there is no straightforward way of deriving the results for one observable from those obtained for the other.

\par  Nevertheless, we should mention that both $T^\text{occ}$ and $T^\text{loc}$ are special cases of another, more general, Brownian functional. Namely, if we choose $V[x(t)]$ to be an indicator function on a segment $[a,b]$, 
\begin{equation}\label{eq:V[x]=indicator general}
  V[x(t)] = \mathds{1}_{[a;b]}[x(t)] \equiv 
  \begin{cases}
      1, & x(t) \in [a;b],\\
      0, & x(t) \notin [a;b],
  \end{cases}
\end{equation}
then occupation time on the half-line and local time density at the origin may be interpreted as the following limits
\begin{align}
  \label{eq:app occ time = def OT1}
  & T^\text{occ} = \lim_{\ell\to\infty}
    \sum_{i} \int_{0}^{t} \mathds{1}_{[0;\ell]}[x_i(t')] \, \dd t' ,
  \\
  \label{eq:app loc time = def OT2}
  & T^\text{loc} = \lim_{\ell\to0} \frac{1}{2\ell} 
    \sum_{i} \int_{0}^{t} \mathds{1}_{[-\ell;\ell]}[x_i(t')] \, \dd t' .
\end{align}
Thus, it would be interesting to study the Brownian functional with the potential \eqref{eq:V[x]=indicator general} and find the behavior of the corresponding probability distributions at large times. Then, by using \eqref{eq:app occ time = def OT1} and \eqref{eq:app loc time = def OT2} we could derive the results for $T^\text{occ}$ and $T^\text{loc}$. However such a computation provide a challenging task, which falls beyond the scope of the present paper.

\par Let us now proceed and compare the results obtained for the occupation time on the positive half-line (which are summarized in Sec.~\ref{seq:results}) with those obtained by the authors previously in \cite{BMR-23} for the local time density at the origin.

\subsection{Typical values}
\par First, let us have a look at the behaviors of the observables close to their typical values. In this case, both quenched and annealed probability distributions are Gaussian, and therefore we can describe them by first two cumulants: the mean and the variance. 
\par The mean values for the quenched and annealed distributions are the same, and for $T^\text{occ}$ and $T^\text{loc}$ they are given by
\begin{equation}
  \overline{ \left\langle T^\text{occ} \right\rangle_\mathbf{x}  } 
   =  \frac{2}{3}\sqrt{\frac{D}{\pi}} 
    \bar{\rho} \, t^{3/2},
  \qquad
  \overline{ \left\langle T^\text{loc} \right\rangle_\mathbf{x} } 
  = \frac{1}{2}\bar{\rho}\,t.
\end{equation}
For the variances in the quenched case we have
\begin{align}
  & \mathrm{Var}_\text{qu}[T^\text{occ}]
  = \frac{8 \left(\sqrt{2}-1\right) }{15} \sqrt{\frac{D}{\pi}}\, \bar{\rho} \, t^{5/2}
  \\
  & \mathrm{Var}_\text{qu}[T^\text{loc}] 
  =
  \frac{2}{3} 
  \left(
    2 - \sqrt{2}
  \right)
  \frac{1}{\sqrt{\pi D}}\, \bar{\rho}\, t^{3/2},
\end{align}  
and the annealed variances at large times behave as
\begin{align}
&\mathrm{Var}_\text{an}[T^\text{occ}] 
  \underset{t\to\infty}{\simeq}
  \frac{2}{5}
  \left[ 1 + \frac{(4\sqrt{2}-7)}{3}(1-\alpha_\text{ic}) \right] \sqrt{\frac{D}{\pi}} \, \bar{\rho} \, t^{5/2},
\\
  &\mathrm{Var}_\text{an}[T^\text{loc}]
  \underset{t\to\infty}{\simeq}
  \frac{2}{3} 
  \left[
    1 + (1 - \sqrt{2})\left( 1 - \alpha_\text{ic} \right)
  \right]
  \frac{1}{\sqrt{\pi D}}\, \bar{\rho}\, t^{3/2}.
\end{align}
\par Note, that the scaling of the mean values and the variances of $T^\text{occ}$ and $T^\text{loc}$ are very different. For the local time density, the mean value grows as $t$ and the variances grow as $t^{3/2}$, whereas for the occupation time the mean value grows as $t^{3/2}$ and the variances grow as $t^{5/2}$. Consequently, for studying the probability distributions in the large deviation formalism, we use different scaling variables, namely
\begin{equation}
  \tau^\text{occ} = \frac{T^{\text{occ}}}{t^{3/2}} \frac{1}{\bar{\rho}\sqrt{4D}},
  \qquad 
  \tau^\text{loc} = \frac{T^{\text{loc}}}{t} \frac{1}{\bar{\rho}}.   
\end{equation}
Let us emphasize, that the observables are very different (they even have different dimensions) and therefore there is no apparent reason for them to have the same scaling. However, there are still some similarities.

\par  First, both for the occupation time and for the local time density, annealed variances depend on the initialization through the single static quantity $\alpha_\text{ic}$ \eqref{eq:alpha_ic=def MainResults}, the Fano factor of the initial condition. Moreover for both observables, the dependence is linear. This is, in fact, expected, since the formalism we used in Sec.~\ref{sec:Mean and Variance: N particles} can be straightforwardly applied to any additive Brownian functional.  

\par 
Second, the ratios between the square of the mean value and the variances are the same for $T^\text{occ}$ and $T^\text{loc}$, namely 
\begin{equation}
  \frac{ \left( \overline{ \left\langle T^\text{occ} \right\rangle_\mathbf{x}  } \right)^2 }
       { \mathrm{Var}_\text{an;qu}[T^\text{occ}] } 
  \underset{t\to\infty}{\sim} 
  \frac{ \left( \overline{ \left\langle T^\text{loc} \right\rangle_\mathbf{x}  } \right)^2 }
       { \mathrm{Var}_\text{an;qu}[T^\text{loc}] } 
  \underset{t\to\infty}{\sim} \bar{\rho} \sqrt{Dt}.
\end{equation}
Hence the large deviation forms are
\begin{align}
  \label{eq:P_anqu=exp[Phi]-occ}
  &\mathbb{P}_\text{an;qu}[T^\text{occ},t] \sim \exp\left[
    -\bar{\rho} \sqrt{4 Dt} \; \Phi_{\text{an;qu}}\left(\tau^\text{occ}\right)
  \right],
  \\
  \label{eq:P_anqu=exp[Phi]-loc}
  &\mathbb{P}_\text{an;qu}[T^\text{loc},t] \sim \exp\left[
    -\bar{\rho} \sqrt{4 Dt} \; \Phi_{\text{an;qu}}\left(\tau^\text{loc}\right)
  \right],
\end{align}
where an additional factor of $2$ in the exponential, is due to purely esthetic reasons. 

\par 
The factor $\bar{\rho}\sqrt{Dt}$ in \eqref{eq:P_anqu=exp[Phi]-occ}, \eqref{eq:P_anqu=exp[Phi]-loc} is again very natural.
The particles do not interact, hence both $T^\text{loc}$ and $T^\text{occ}$ are essentially sums of $N$ independent random variables. However, a lot of these random variables are zeros. 
Indeed, only the particles that have reached the origin up to the observation time have nonzero contribution to both $T^\text{occ}$ and $T^\text{loc}$. Since the  typical displacement  of a Brownian particle is proportional to $\sqrt{Dt}$, we expect, that the effective number of particles is proportional to $\bar{\rho}\sqrt{Dt}$. 
In other words, we have effectively $n \sim \bar{\rho}\sqrt{Dt}$ terms in \eqref{eq:app occ time = def} and \eqref{eq:app loc time = def}, and therefore it is natural to expect this sums to be proportional to $n$. However, we should stress, that since we are dealing with nonidentically distributed random variables, the above argument is somewhat handwaving.

\subsection{Atypical values}
\par
\begin{figure}
\includegraphics[width = .8\linewidth]{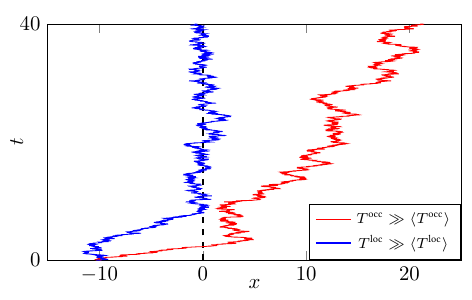}
\caption{Examples of the Brownian trajectories with atypically large value of the occupation time (red) and local time density at the origin (blue). These two  trajectories are qualitatively different. In the simulation $t=40$, $x(0)=-10$, and $D=1$. To obtain this picture, we have generated $10^7$ independent Brownian trajectories and picked two of them corresponding to the largest values of $T^\text{loc}$ and $T^\text{occ}$.}
\label{fig:example OT and LT}
\end{figure}

 Now, let us comment on the tails of the probability distributions of $T^\text{occ}$ and $T^\text{loc}$. Note, that if one observable is equal to zero, then the other one equals zero as well. This is simply due to the fact, that both situations correspond to configurations of the trajectories in which no trajectory crosses the origin. Therefore we expect that $\mathbb{P}[T^\text{loc}=0,t] \sim \mathbb{P}[T^\text{occ}=0,t]$. This is in perfect agreement with our results, 
\begin{align}
  & \mathbb{P}_\text{an}[ T^\text{occ}=0,t]
  \sim
  \mathbb{P}_\text{an}[ T^\text{loc}=0,t] \sim  e^{ - \theta_\text{an} \, \bar{\rho} \sqrt{4Dt}  }
  \\
  & \mathbb{P}_\text{qu}[ T^\text{occ}=0,t]
  \sim
  \mathbb{P}_\text{qu}[ T^\text{loc}=0,t] \sim  e^{ - \theta_\text{qu} \, \bar{\rho} \sqrt{4Dt}  }
\end{align}
As for the atypically large values, the situation is different. Recall, that for the occupation time we have
\begin{align}
    & \mathbb{P}_\text{an}[T^\text{occ},t] 
     \underset{T^\text{occ} \to\infty}{\sim} 
    \exp\left[ - {\textstyle \frac{T^\text{occ}}{2t}} 
      \log\left( {\textstyle \frac{\left(T^\text{occ}\right)^2}
                                  {t^3D\bar{\rho}^2}}\,  \right) \right],
    \\
    & \mathbb{P}_\text{qu}[T^\text{occ},t] 
    \underset{T^\text{occ} \to\infty}{\sim} 
    \exp\left[ - {\textstyle \frac{16}{27}\frac{T^\text{occ}}{t}} 
      \left( {\textstyle\frac{\left(T^\text{occ}\right)^2}{t^3D\bar{\rho}^2}}\right) \right].
\end{align}  
whereas for the local time density
\begin{align}\
  \label{eq:P_an_tail=}
  &
  \mathbb{P}_\text{an}[T^\text{loc},t] 
  \underset{T^\text{loc} \to\infty}{\sim} 
  \exp\left[
    - \sqrt{D}\,  \textstyle\frac{T^\text{loc}}{\sqrt{t}}  \sqrt{ \log \frac{T^\text{loc}}{t\bar{\rho} } } 
  \right]
  \\
  \label{eq:P_qu_tail=}
  & 
  \mathbb{P}_\text{qu}[T^\text{loc},t] 
     \underset{T^\text{loc} \to\infty}{\sim}  
    \exp \left[ 
      - 
      \sqrt{\frac{32}{27}} \,
      \sqrt{D}
      \frac{ \left( T^\text{loc} \right)^{3/2}}{t\sqrt{\bar{\rho}}}  
      \right],
\end{align}
There seems to be no similarity between the $T^{\text{loc}}$ and $T^{\text{occ}}$ but this should not surprise the reader. The trajectories, that contribute to the atypically large values of these observables are qualitatively different. 
This becomes clear when considering a single particle problem. 
Trajectories with $T^\text{loc}\gg\left\langle T^\text{loc}\right\rangle$ tend to remain close to the origin most of the time. On the other hand, trajectories with $T^{\text{occ}}\gg\left\langle T^{\text{occ}}\right\rangle$ can go arbitrary far from the origin (see Fig.~\ref{fig:example OT and LT}).

\end{appendix}

\bibliography{OT_BM.bib}

\end{document}